\documentclass[aps,12pt,showpacs]{revtex4}

\usepackage[dvips]{graphicx} 
\usepackage{latexsym}        
\usepackage{amsfonts}        
\usepackage{amssymb}
\usepackage{amsmath}
\usepackage{longtable}
\usepackage{flafter}
\usepackage{bm}              
\usepackage{epsfig}
\usepackage{float}

\usepackage{times}           

\newcommand{\qq}{\mbox{$\mathbf{q}$}}

\newcommand{\rr}{\mbox{$\mathbf{r}$}}
\newcommand{\vv}{\mbox{$\mathbf{v}$}}

\begin{document}

\title{Super-heterodyne light scattering on interacting colloidal suspensions: theory and experiment}

\author{Thomas Palberg, Holger Reiber, Tetyana K\"oller}
\affiliation{Johannes Gutenberg Universit\"at, Institut f\"ur Physik,
Staudingerweg 7, D-55128 Mainz, Germany}
\author{Martin Medebach}
\affiliation{Technische Universit\"at Berlin, Ivan-N.-Stranski Laboratory,
Strasse des 17. Juni 135, D-10623 Berlin, Germany}
\author{Gerhard N\"agele}
\affiliation{Forschungszentrum J\"ulich GmbH, IFF, D-52425 J\"ulich, Germany}

\begin{abstract}
In soft matter structure couples to flow and vice versa. Complementary to structural investigations, we here are interested in
the determination of particle velocities of charged colloidal suspensions of different structure under flow. In a combined
effort of theory and experiment we determine the Fourier transform of the super-heterodyne field auto-correlation function
(power spectrum) which in frequency space is found to be well separated from homodyne contributions and low frequency noise.
Under certain conditions the power spectrum is dominated by incoherently scattered light, originating from the unavoidable size
polydispersity of colloidal particles. A simple approximate form for the low-wavenumber self-intermediate scattering function is
proposed, reminiscent to the case of non-interacting particles. We experimentally scrutinize the range of applicability of these
simplified calculations on employing a parabolic electro-osmotic flow profile. Both for non-interacting and strongly interacting
fluid particle systems, the spectra are well described as diffusion-broadened velocity distributions comprising an osmotic
flow-averaged superposition of Lorentzians at distinct locations. We discuss the performance and scope of this approach with
particular focus on moderately strong interactions and on multiphase flow. In addition, we point to some remaining theoretical
challenges in connection to the observed linear increase of the effective diffusion constant and the integrated spectral power
with increasing electric field strength.
\end{abstract}


\date{{\today}}  

\maketitle

\section{Introduction}

Soft condensed matter is typically characterized both by its softness and an internal structure on a mesoscopic scale. This
allows convenient optical access to system structure and dynamics in equilibrium. Structural properties, obtained from static
light scattering with good statistical accuracy can be compared to theoretical expectations and/or computer simulations
\cite{Pusey,Nagele:96,LöwenHabil}. In particular for systems of strongly interacting particles, local and global changes of
structure and even phase transitions are frequently observed with and without flow, thus providing an interesting field for
studies of non-equilibrium phenomena \cite{Ackersonshearrev,chen discont shear thinning, Berret Wormlike micelles,Dhont shear
banding,Biehl EPL, Ackerson Phase transitions,Löwen Rev noneq.}. We are intererested here with the complementary task of
determining the (local) particle velocities in systems of interacting particles in arbitrary flow profiles (plate-plate or cone
plate shear, Hagen-Poiseuille flow, electrophoretic-electro-osmotic flow in closed cells, thermophoresis, sedimentation etc.)
and different structure (unstructured, fluid, crystalline) The combination of structure and flow measurements in suitable model
systems should in the long run help to elucidate the fundamental coupling between both.

A well developed arsenal of methods is available to study the flow
behavior of fluids in general. Quite frequently tracer particles are
used \cite{Adrian Ann rev fluid mech}. Direct particle imaging
nowadays is capable of fast scans of 3D velocity distributions,
e.g., in gas turbulence \cite{Atlanta} or wide field scans with high
spatial resolution \cite{Barnhart} (in particular holographic image
velocimetry displays diffraction limited spatial resolution
\cite{Adrian Ann rev fluid mech,Sherer}). Both imaging applications,
however, often require a complex instrumentation, even if for 2D
scans the less demanding speckle imaging velocimetry can be used
\cite{Alaimo}. Tracer experiments have also been reported on
suspensions of interacting particles, but in general the choice of
suitable systems is strongly restricted \cite{vanMegen tracer}.
Laser Doppler velocimetry (LDV) on the other hand, which is a
heterodyne light scattering technique, can be directly employed with
pure samples and small particle sizes. LDV uses either real fringe
or reference beam realizations \cite{Adrian selected papers}. Using
fringe techniques, two laser beams are crossed in the sample cell to
produce a local illuminating grating. For sufficiently dilute
suspensions the particle motion can then be traced via an
oscillating scattered intensity of frequency proportional to the
particle velocity. For large particle concentrations the analysis of
the scattered intensity yields the heterodyne intensity
autocorrelation function (IACF) or its Fourier transform, the power
spectrum. Contributions of low frequency noise
and the homodyne IACF (stemming from the self beat of photons
scattered off the sample) as well as the overlap of the two Fourier
transform components of the heterodyne IACF \cite{TPHV} (observed
for low velocities) can be avoided by introducing a frequency shift
between the two illuminating beams. Frequency shifting was first
used for electrokinetic applications by Sch\"atzel et al. to locally
separate the diffusion and the small drift velocities in
non-interacting particle systems \cite{Miller Schätzel EOflow}. It
may be noted that in that work also an alternative data analysis was
employed to obtain the so-called amplitude weighted phase structure
function from the phase of the scattered light \cite{SchätzelMerz}.
The main general drawback of fringe techniques however is their
strictly local velocity detection implying a tedious point-by-point
determination of the actual flow pattern \cite{TPMW multiphase
flow}.

By contrast in a reference beam set-up, a small portion of the
illuminating beam is separated. It is re-combined at the detector
place with the Doppler-shifted light scattered off a small volume of
the sample to act there as local oscillator and give rise to beats
in the detected intensity \cite{Ware:74,Berne Pecora Book,Dhont
book}. The beat frequency is proportional to the Doppler frequency
and therefore can be analyzed by considering either the IACF or its
Fourier transform pair, the power spectrum. As the
illuminating beam crosses the complete sample cell (integral
measurement \cite{TPHV}), in principle one obtains information on
the velocity distribution in a non-interacting sample, provided the
scattering particles are homogeneously distributed and all of same
scattering power. Then the amplitude of the spectrum contains
information on the number of scatterers moving with a given
velocity, and for a known sample cell geometry the flow profile can
be quantified.

The situation of interacting samples, where a locally different
structure will lead to different amplitudes has not been rigorously
treated. The present paper will make a first step in that
direction. Further, in previous studies, the reference beam setup
suffered from the restriction that the homodyne contribution
centered around zero frequency could possibly overlap with the
heterodyne scattering contribution containing the information on the
particle velocities. In the present work, we use a straightforward
extension of the approach, used by Sch\"atzel et al. for fringe
techniques, to integral reference beam techniques by introducing a
frequency shift between illuminating and reference beam. This shift
is also accounted for by us in the calculation and analysis of the
super-heterodyne signal. The general theory of dynamic light
scattering from colloidal particles is known since many years and
has been summarized, e.g., in \cite{Berne Pecora Book,Dhont
book,Cummins:69}. The heterodyne (electrophoretic) light scattering
technique for dilute systems of non-interacting particles has been
discussed thoroughly in \cite{Ware:74}. The shifting method has been
very shortly discussed in \cite{Miller:92}. The present paper gives
a theoretical derivation of the super-heterodyne intensity
autocorrelation function and its Fourier-transform pair, the power
spectrum.

Moreover, this derivation for the first time includes an account of
the unavoidable polydispersity actually leading to a principal
possibility to deal with differently structured systems. To be
specific, the theoretical analysis of the super-heterodyne IACF
reveals that the shape of the measured power spectrum is usually connected to a
convolution of the particle velocity distribution and the
microstructure of particles, since the heterodyne part of the IACF
contains an amplitude factor related to the local 'static' structure
factor. Thus, for suspensions of correlated particles, the
assumption of a homogeneous distribution of scattering power may be
invalidated if the microstructure of the system changes locally. It
is a main point of the present paper to experimentally realize the
solution to this problem as suggested by theory. Exploiting the
dominance of incoherently scattered light at low scattering angles
and strong particle interactions, we show that the spectra obtained
from electro-osmotically sheared suspensions of fluid-like order
still can be unequivocally interpreted as velocity distributions.

In addition  we also observe an increase of the effective
diffusion coefficient and the integrated spectral power with the
electric field for the case of suspensions of interacting particles.
The field-dependence of both quantities is approximately linear in
the case of fluid-ordered systems under stationary flow, but more
complex in the case colloidal crystals. Interestingly, the
integrated power stays constant during shear-induced melting at
constant applied field, suggesting that the additional spectral
power is not induced by the structural change.

While the physics behind these two effects are not yet fully
understood, our tests clearly show that we have a versatile tool for
quantifying the flow behavior in the fluid-ordered state, and for
qualitatively complementing structural measurements in the
crystalline state and in multi-phase flow. Given this, our approach
should have several interesting applications. First, it could be
quite useful to characterize deviations from simple Hagen-Poiseuille
flow in non-Newtonian fluids \cite{Ackersonshearrev,chen discont
shear thinning,Berret Wormlike micelles,Dhont shear banding, Biehl
EPL,Ackerson Phase transitions,Löwen Rev noneq.,
SchätzelMerz,Bergenholtz HI shear thinning,TPSvHMW multiphase
flow,Preis flow profiles}. From a comparison of experimental
results, it could be useful to separate the effects of shear,
absolute drift velocity against the solvent and external fields on
the suspension dynamics, thus enabling an extension of presently
available theoretical approaches, e.g., to investigations on the
dynamic structure factor for interacting suspensions under flow
\cite{Dhont book}. Our method should be also well suited to study
laning effects occurring in mixtures with different particle speeds
\cite{Dzubiella PRE laning}. And finally, it has already turned out
to be useful for quantitative electrophoresis experiments in
interacting suspensions \cite{MM CSA 2003,MM JCP 2003,MM JPCM
2004,MM CSB 2007,ReiberKöller Ludox}.

In what follows, we shall first explain the experimental
set-up for integral super-heterodyne light scattering in the
reference beam mode, then give a thorough derivation of the IACF and
the power spectrum for moderately polydisperse and strongly
interacting charged-sphere suspensions. We then show that the
assumptions made in our theoretical derivations can be well met in a typical
experiment and verify the theoretical results with experiments on
different experimental situations. We finally discuss the
performance and scope of the super-heterodyne velocimetry method and give an outlook on open
theoretical and experimental issues.

\section{Experimental}
\subsection{Integral Super-heterodyne dynamic light scattering set-up}

\begin{figure} \vspace*{2cm}
\epsfig{file=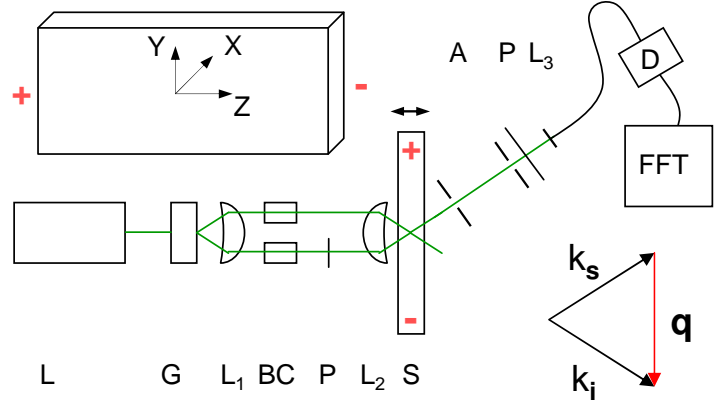,width=9.5cm,angle=0} \caption{\label{fig:1}
Super-heterodyne velocimetry set-up. Top: Sketch of the rectangular
flow-through cell and the used coordinate system. The plus and minus
signs indicate the electrodes used to induce the combined
electrophoretic-electro-osmotic flow. Middle: Top view of the
super-heterodyne experiment (for a detailed description see text).
Bottom right: definition of the scattering wave vector ${\bf q}$.}
\end{figure}

Colloidal particle velocities have been obtained from a home-built
super-heterodyne reference beam mode Laser Doppler velocimetry setup. Our
instrument is a consequent extension of the heterodyne reference
beam technique presented previously \cite{TPHV}, but in contrast to
the latter contains a frequency shift between illuminating beam and
reference beam. This ensures excellent discrimination of the
electrophoretic signal from all other spectral components and low-frequency
electronic noise. The set-up is shown schematically in
Fig. \ref{fig:1}. The beam of a Nd-Yag laser ($L$, $\lambda_{0} = 532\;\!\mathrm{nm}$) of
circular frequency $\omega_0 \sim 10^{15}$ Hz is split into two
beams, a reference beam (r) and an illuminating beam (i), using a beam
splitter. Alternatively also a sinusoidal transmission grid ($G$)
can be used (Sine Patterns, Pittsford NY). The two beams are made parallel by a lens $L_{1}$ and each beam
passes through a Bragg cell ($BC$) and is frequency shifted by
the circular frequencies $\omega_i = 1.002$ MHz and $\omega_r = 1$ MHz, respectively, with a
positive relative Bragg shift, $\Delta \omega_B = \omega_i
-\omega_r >0$, in the circular frequency corresponding to $\Delta f_{B} = \Delta\omega_{B}/
2\pi  = (2000\pm 5)$
Hz. The larger frequency is realized for
the illumination beam. The beams are redirected into the sample cell
($S$) by lens $L_{2}$ where they cross in suspension under an angle
of $\Theta =18.82°$.

In addition to a polarizer ($P$) the detector side optics include a
a set of apertures ($A$) to define the observed sample volume and an
optical fibre to define the scattering vector ${\bf q}$.  The
scattering vector is
\begin{equation}
\label{eq:scattering-vector}
   {\bf q} \;\!=\;\! {\bf k}_i - {\bf k}_s
\end{equation}
where ${\bf k}_i$ and ${\bf k}_s$
are the wave vectors of the illuminating and scattered,
respectively (c.f., Fig. \ref{fig:1}). The modulus of the scattering vector is
given as $q = (4\pi \nu_{s}/\lambda_{0})\sin(\Theta/2) = 5.0\;\!\mu
\mathrm{m}^{-1}$, where $\nu_{s}$ the index of refraction of the solvent.
For the present sign convention, $\qq$ is proportional to the momentum transfer
from the photon to the scattering particle.
For so-called integral measurements, light scattered off the
illuminated part of the sample is received within the acceptance
angle of a grin lens. This lens is mounted at the inlet of an optical fibre
leading to a photomultiplier used to record the intensity ($D$). Note
that for the present definition of $\qq$, the light scattered by a particle $j$ moving with a velocity
${\bf v}_{j}$ is Doppler-shifted by the circular frequency $\omega_\mathrm{D} = - {\bf
q}\cdot{\bf v}_{j}(x,y)$, where the velocity could be a
function of the position. The frequency shift $\omega_\mathrm{D}$ is positive, if the particle
velocity shares an obtuse angle with the scattering vector $\qq = q\;\! \widehat{\bf z}$, where
$\widehat{\bf z}$ is the unit vector in $z$-direction. For example, in Fig. \ref{fig:1} a
positive Doppler shift corresponds to particles moving towards the
detector. Also the reference beam is directed into the fibre and
superimposed with the scattered light. It therefore acts as a local
oscillator, and gives rise to beats in the observed intensity which
are analyzed by a Fast Fourier Transform analyzer (Ono Sokki DS2000,
Compumess, Germany) to yield the power spectrum as a function of
frequency $f = \omega/2\pi $. As we will discuss, the spectrum
is composed of a strong peak at zero frequency, a
homodyne term stemming from the self-beating of the scattered light,
which is localized  around zero frequency, and a
super-heterodyne part, which is Doppler-shifted relative to the Bragg frequency shift
$ \Delta f_B$. The latter term contains all the information on
particle velocity and direction and is further evaluated.

\begin{figure} \vspace*{0cm}
\epsfig{file=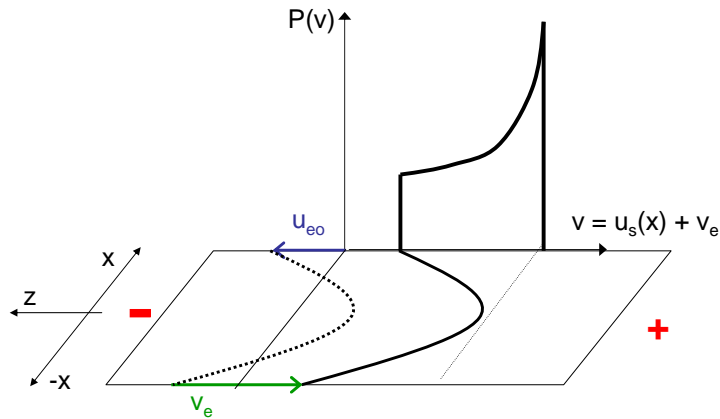,width=9.5cm,angle=0} \caption{\label{fig:2}
Horizontal $x-z$ midplane: laboratory-frame velocity profiles
$v(x)=v(x,y=0)$ and $u_S(x)=u_S(x,y=0)$ of particles and solvent,
respectively. Vertical plane: corresponding probability density,
$P(v)$, of finding a particle of given velocity $v$. In our setup,
the external field ${\bf E}$, ${\bf u}_{eo}$ and $\qq$ are aligned
with the positive $z$-axis, whereas $\vv_e$ points into the negative
$z$-direction.}
\end{figure}

In our measurements, we have opted for a parabolic solvent flow profile.
Such a profile can be easily realized in a cylindrical tube, through which a
suspension flows under a hydrostatic pressure difference.
For strongly interacting suspensions, however, this
approach often suffers from an insufficient stationarity of the flow,
and an only moderate reproducibility \cite{Preis flow profiles}.
Therefore, in the experiments discussed here, we have applied
an electric field to move the particles and shear the suspension.
Application of an external static electric field, ${\bf E}$, pointing
into the positive $z$-axis in a closed cell geometry (see Fig. \ref{fig:1})
allows for the development of an electro-osmotic flow profile of
the incompressible solvent. Surface-released cations in the thin double layer at the
negatively charged cell walls move towards the negative electrode
with a velocity $u_{eo} > 0$ pointing along the positive $z$-axis (see Fig. \ref{fig:1})
and drag some solvent along with them. A central solvent backflow
in direction of the negative $z$-axis assures volume
conservation of the incompressible fluid (Fig. \ref{fig:2}). For DC electric fields a stationary parabolic flow
profile results in both cylindrical geometry and the mid-plane of a
rectangular cell \cite{Komagata}. For optical reasons we here
employ a rectangular cell as shown with its electrodes in Fig. \ref{fig:1}.
The electro-osmotic flow profile, $u_S(x)$, of the solvent is sketched in Fig. \ref{fig:2}
for the $x-z$ mid-plane located  mid-cell of the rectangular cell at $y=0$.
Note that $u_S(x)$ is equal to $u_{eo}$
at the side walls, and due to the rectangular cell shape, the integral of $u_S(x)$
with respect to $x$ from one side wall to the other
is zero due to volume conservation of the incompressible solvent.

Superimposed on this solvent flow profile is the electrophoretic
motion of charged colloidal particles relative to the solvent with the electrophoretic
velocity $v_{e} = \mu_e \;\!E$, where $\mu_e$ is the electrophoretic mobility.
The constant electrophoretic velocity
$v_{e} < 0$ of the negatively charged colloids points in negative $z$ direction towards the positive
electrode. The resulting flow
profile,
\begin{equation}
\label{particle-velocity-profile}
   v(x) \;\!=\;\! v_{e} + u_S(x)  \,,
\end{equation}
of colloidal particles in the laboratory frame is thus equal to the solvent profile
shifted by the constant amount $|v_{e}|$ in negative $z$ direction. Also drawn schematically in the Fig. \ref{fig:2}
is the particle velocity distribution, $P(v)$, where $P(v)\;\!dv$ is the fraction of colloids
with velocity in a infinitesimal interval $dv$ around $v$.  On assuming a homogeneous
distribution of colloidal particles that remains unperturbed by the electro-osmotic flow,
$P(v) \propto dx/dv$ \cite{TPHV}.

Clearly the suspension is sheared at locally
different rates by the underlying solvent flow. All
colloidal particles and their electric double layers are subjected to the homogeneous
electric field that couples differently to colloids and microions.
If instead shear motion is
introduced by a hydrostatic pressure difference and not by an electric field, the microions
and the colloids are affected in the same way by the flow.
If sedimentation is addressed in place of electrophoresis,
gravity couples to the colloidal
particles but not to the microions in the double layers and the solvent flow in sedimentation is
not parabolic. Pipe flows are just one example of many possible
shear profiles. For interacting colloidal particles, shear flow may lead to local
phase transitions and, in concert with hydrodynamic interactions (HI), to
a non-Newtonian flow behavior. This has been observed for both
pressure and electro-kinetically driven flows
\cite{Ackersonshearrev,chen discont shear thinning,Berret Wormlike micelles,
Dhont shear banding,Biehl EPL,TPMW multiphase flow,TPSvHMW multiphase flow,
Preis flow profiles,MM CSA 2003,MM JCP 2003,MM JPCM
2004,MM CSB 2007}. A fascinating richness of structural effects was
there observed ranging from (partial) shear melting, to shear
banding and shear induced structure formation, which all coupled
back to the flow properties. In practically all cases deviations
from the parabolic flow profile were observed. For the present paper
we concentrate on one-phase systems of more or less pronounced fluid-like
order and only shortly strive suspension flows with phase transitions.
Consequently the theoretical section \ref{sec:Theory} is concerned with the
derivation of the IACF and the corresponding super-heterodyne
spectrum for homogeneous systems in a stationary state.

\subsection{Samples and sample preparation}

\begin{figure} \vspace*{0cm}
\epsfig{file=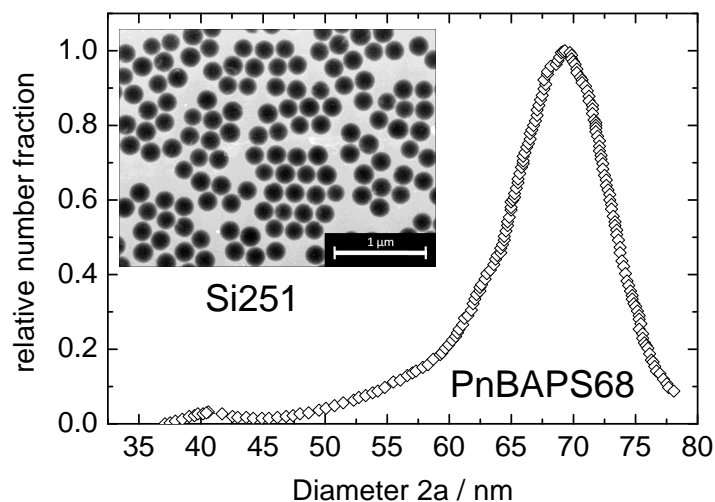,width=9.5cm,angle=0} \caption{\label{fig:3}
Particle size distribution from ultracentrifugation for PnBAPS68.
The insert shows a TEM picture of Si251. The scale bar is $1 \;\!\mu
\mathrm{m}$.}
\end{figure}

One species of colloidal particles employed in this work are silica spheres
synthesized by a modified St\"ober synthesis, showing a diameter of
$2a=251$ nm and a relative standard deviation in the size distribution of $s = 0.08$
determined from TEM-measurements. A representative TEM image is shown in the insert
of Fig. \ref{fig:3}. Suspensions of Si251 particles could be used in our measurements only
in the fluid-like phase. Working at larger particle concentrations was inhibited by the onset
of strong multiple scattering. The second species of spherical particles we have made use of is an industrial
sample of Poly-n-Butylacrylamide-Polystyrene copolymer and a kind
gift of BASF, Ludwigshafen. The PnBAPS68 particles have a diameter of $2a =
68$ nm and a standard deviation of $s = 0.05$ (determined from ultracentrifugation)
and carry an effective charge of $Z = 420\;\!\mathrm{e}^{-}$ obtained from
conductance measurements. The particle size distribution deduced from
ultracentrifugation is shown in Fig. \ref{fig:3}. Aqueous suspensions of this species can be
investigated up to large particle concentrations without too much
multiple scattering, so that we have studied them also in the
crystalline state.

Both particle systems were prepared from pre-cleaned stock suspensions of
large particle concentration stored over a mixed-bed ion exchanger. By
precise dilution with milli-q-grade water, and by determining the particle
density, $n$, using static light scattering,  the desired number densities
have been adjusted and the resulting suspensions filled in a closed
Teflon®-tubing circuit connecting an ion exchange chamber, a
reservoir under an inert $Ar$ atmosphere (to add further suspension,
salt or solvent), and a set of different measurement cells. These
comprised a conductivity experiment, a cell for static light
scattering, and the flow-through cell for the velocity measurements.
Details of the used preparation circle are given elsewhere
\cite{Continuous deionization,continuous deionization improved}. We
here only note that residual impurity concentrations on the level of
the self-dissociation product of water are reached within a few
hours. For measurements at elevated salt concentrations, $NaCl$
solution was added after complete deionization. Continued cycling
under bypassing the ion exchange column ensured homogenization, on
controlling the salt concentration by conductivity measurements \cite{independent
ion migration} and the particle concentration by static light
scattering with a residual uncertainty of about one percent. This
preparation method is thus very fast and assures good
reproducibility of adjusted suspension parameters. Furthermore, the
preparation circuit permits to conduct several experiments
simultaneously. This is of great importance to have a continuous control
on the constancy of interaction parameters.

\subsection{Electrokinetic flow experiment}
Electrophoretic-electro-osmotic shear flow experiments were
performed in closed quartz cells of rectangular cross section with
depth $\times$ height $= d \times  h = 1\;\! \mathrm{mm} \times 10
\;\! \mathrm{mm}$ (Rank Bros. Bottisham Cambridge, UK). As sketched
in Fig. \ref{fig:1} on the upper left, the origin of our coordinate
system is at the cell center with $x\parallel d, y\parallel  h$ and
$z\parallel l$. The optical part of the cell length, $l$, spans some
$40$ mm and the effective platinum electrode distance is $l = 85.5$
mm as determined from calibration with an electrolyte dilution
series. During the measurements the electrode chambers are sealed
against the remaining preparation circuit by electromagnetic valves
closed upon stop of pumping. To avoid accumulation of particles at
the electrodes alternating square-wave fields of strength up to
$E_{max} = U/l =  80$ V/cm were applied to fluid ordered systems. To
ensure fully developed stationary flows, field switching frequencies
$f_{AC} = (0.05 - 0.2)$ Hz were used. Measurement intervals were
restricted to one field direction, starting 2s after field reversal
and ending 1s before the next field reversal.

\section{THEORY } \label{sec:Theory}
In this section we give the essentials of the theory of
super-heterodyne electrophoretic light scattering from suspensions
of charge-stabilized colloidal spheres subject to a constant
external electric field. We proceed from
non-interacting particles to low-salinity
suspensions of strongly interacting particles. To our knowledge,
the super-heterodyne electrophoretic scattering theory for
strongly interacting colloidal particles has not been presented so far.

For a stationary and spatially homogeneous system, we
derive the general expression for the measured intensity
autocorrelation function (IACF) and its related power spectrum (PS)
in terms of (partial) dynamic structure factors. The structure
factors characterize the density fluctuation correlations of
colloidal spheres in the presence of an external electric field. It
will be discussed on how the electrophoretic velocity of the spheres may be
deduced from small wave number measurements on weakly
polydisperse systems.

\subsection{Super-heterodyne scattering theory}
\label{subsec:scattering_theory}

The super-heterodyne scattering setup has been shown schematically
in Fig. \ref{fig:1}. We assume in the following that the colloidal
spheres are optically isotropic, and that
the Rayleigh-Gans-Debye description of weakly scattering particles
applies where the electric fields
scattered at each point inside a sphere can be linearly
superimposed. The following general discussion does not include electro-osmotic
flow effects which are of importance in closed cell geometries. The influence
of a non-zero macroscopic solvent flow profile on the measured super-heterodyne power spectrum
will be discussed in subsection \ref{solvent-flow-profile}.

Let the Bragg-shifted illuminating electric field be a plane wave of the form
\begin{equation}
\label{eq:Eincident}
    {\bf E}_i({\bf r},\tau) \;\!=\;\! \hat{\bf n}_i \;\! E_i^0 \;\! \exp\{i\;\! \left( {\bf k}_i \cdot {\bf r} -
    \left[ \omega_0 + \omega_i \right]\tau \right) \} \,,
\end{equation}
where $\hat{\bf n}_i$ is the unit polarization vector of the incident field.
Under the assumptions made above, and for the standard V-V
geometry of polarized scattering, the field strength,
$E_s(\qq,\tau)$, of single-scattered light from $N$ spheres in
the scattering volume follows from standard scattering theory as \cite{Berne Pecora Book,Dhont book,Chu-book}
\begin{equation}
\label{eq:Escattered}
    E_s(\qq,\tau) \;\!=\;\! \exp\{-i\;\!(\omega_0 + \omega_i)\;\! \tau \} \;\!c\;\!
                    \sum_{l=1}^N f_l(q) \;\! \exp\{i \qq \cdot \rr_l(\tau) \}
                    \,,
\end{equation}
with ${\bf q}$ as defined in Eq. (\ref{eq:scattering-vector}). Here,
$f_l(q)$, is the scattering amplitude of a sphere labelled by $l$,
whose center at time $\tau$ is at the position $\rr_l(\tau)$. For a
homogeneous sphere of radius $a_l$, volume $V_l$ and refractive
index $\nu_l$, the scattering amplitude is given by $f_l(q)=
2\;\!\!(\nu_l - \nu_s)\;\!\!V_l\;\! b_l(q)/\nu_s$, with the form
amplitude $b_l(q) = 3 j_1(qa_l)/(qa_l)$, where $j_1$ is the
spherical Bessel function of first order. The apparative constant
\begin{equation}
\label{eq:constant-c}
    c \;\!=\;\! E_i^0 \; \frac{\exp\{i\;\!|{\bf k}_s| \;\!R\}}{\lambda^2\;\!R}
    \end{equation}
depends on $\lambda$, $k_s$, the large sample-to-detector distance
$R$, and the field amplitude, $E_i^0$, of the illuminating light
\cite{Dhont book,Berne Pecora Book}. Its explicit form is of no
concern, constant throughout the experiments and thus will not be
displayed any more. The corresponding electric field strength of the
reference beam at the detector is
\begin{equation}
\label{eq:Ereference}
    E_r(\tau) \;\!=\;\! \;\! E_r^0 \exp\{-i\;\!(\omega_0 + \omega_r)\;\! \tau
    \} \,.
\end{equation}
The reference field is independent of the scattering wave vector and
the particle configuration.

In the (super-)heterodyne setup, the intensity autocorrelation
function of the mixed electric fields is analyzed by a real-time
auto-correlator. For a stationary system, the
super-heterodyne mixed-field intensity autocorrelation function (IACF) is given by
\begin{equation}
\label{eq:IACFmixed}
    C_\textrm{shet}(\qq,\tau) \;\!=\;\! \;\! \left\langle \;{|E_s(\qq,\tau) + E_r(\tau)|}^2
                   \cdot {|E_s(\qq,0) + E_r(0)|}^2 \; \right\rangle \,,
\end{equation}
where $\tau$ is the correlation time, which in our experiments is
$10 - 100$ ms or larger. The angular brackets
denote a time average or likewise, for a stationary and ergodic system, an
ensemble average. The IACF
depends on the orientation of $\qq$,
since spatial isotropy is broken by the
applied electric field. Carrying out the multiplications gives
sixteen terms, ten of which vanish due to the assumed spatial homogeneity,
and three of which are time-independent. The
remaining three terms determine the time dependence of the IACF. To
see this, we rewrite Eq. (\ref{eq:IACFmixed}) as
\begin{eqnarray}
\label{eq:IACF-multiplied}
    C_\textrm{shet}(\qq,\tau) &=& \left\langle
    \left[I_s(\qq,\tau) + I_r(\tau) + E_r(\tau) E_s^\ast(\qq,\tau) +  E_r^\ast(\tau) E_s(\qq,\tau)
    \right] \right.
    \times \nonumber \\
    & & \left. \;\; \left[ I_s(\qq,0) + I_r(0) + E_r(0) E_s^\ast(\qq,0) +  E_r^\ast(0) E_s(\qq,0)\right]  \right\rangle \,,
\end{eqnarray}
with $I_s(\qq,\tau)= E_s^\ast(\qq,\tau) \cdot E_s(\qq,\tau)$ and
$I_r(\tau)= E_r^\ast(\tau) \cdot E_r(\tau)$. The asterisk denotes the operation of complex
conjugation. The reference beam
quantities, $I_r$ and $E_r$, are independent of the ensemble average
over the particle configurations. Moreover, the averages over
products of an odd number of factors $E_s(\qq)$ and/or
$E_s^\ast(\qq)$ vanish for $q>0$, since spatial homogeneity requires
the wave vector sum to be zero. For the same reason,
$\left\langle E_s(\qq,\tau) E_s(\qq,0) \right\rangle$ and
$\left\langle E_s^\ast(\qq,\tau) E_s^\ast(\qq,0) \right\rangle$ are
identically zero. Using these properties of spatial homogeneity, we
multiply the first element, $I_s(\qq,\tau)$, in the upper bracket of
Eq. (\ref{eq:IACF-multiplied}) with the elements of the lower
bracket, keeping terms only which are non-zero after averaging. On
repeating this multiplication for the remaining three elements in
the upper bracket, we obtain the
intermediate result
\begin{eqnarray}
\label{eq:IACF-result}
    C_\textrm{shet}(\qq,\tau) &=& I_r \left[\; I_r + 2 \left\langle
    I_s(\qq) \right\rangle \; \right] + 2 \;\! \textrm{Re} \left\{\;\!  E_r^\ast(\tau)\;\! E_r(0)\;\!
    \left\langle E_s(\qq,\tau) \cdot E_s^\ast(\qq,0) \right\rangle\;\!\right\}
    + \nonumber \\
     & & \, \left\langle {|E_s(\qq,\tau)|}^2 \cdot {|E_s(\qq,0)|}^2
    \right\rangle  \,,
\end{eqnarray}
showing that three out of six non-zero terms are time-independent,
and add up to a constant baseline $I_r \left[ I_r + 2 \left\langle
I_s(\qq) \right\rangle \right]$, where $I_r = {|E_r^0|}^2$
and $I_s(\qq) = {|E_s(\qq,0)|}^2$. To make the frequency dependence
of $C_\textrm{shet}(\qq,\tau)$ explicit, we redefine $E_s(\qq,\tau)$ now
as
\begin{equation}
\label{eq:Escattered-redefined}
    E_s(\qq,\tau) \;\!=\;\! c\;\!
                    \sum_{l=1}^N f_l(q) \;\! \exp\{i \qq \cdot \rr_l(\tau)
                    \} \,.
\end{equation}
With this redefinition, we obtain
\begin{eqnarray}
\label{eq:IACFmixed-final}
    C_\textrm{shet}(\qq,\tau) = I_r \left[ I_r + 2 \left\langle
    I_s(\qq) \right\rangle \right]+ 2\;\!I_r\;\! \textrm{Re} \left[\;\!g_E(\qq,\tau)
    \exp\{- \;\!i\;\! \Delta\omega_B\;\! \tau \}\;\!\right] +
    g_I(\qq,\tau) \,,
\end{eqnarray}
for the super-heterodyne scattering function, where $\Delta \omega_B = \omega_i
-\omega_r > 0$. The standard expression
for heterodyne scattering is recovered for zero Bragg shift \cite{Cummins:69}.
Note that $C_\textrm{shet}(\qq,\tau)$ does not depend on $\omega_0$ since,
due to the mixing of $E_s(\qq,\tau)$ and $E_r^\ast(\tau)
$, it only contains the
low-frequency beats.

Information on the particle motion is embedded in the
super-heterodyne part of the mixed-field IACF, given by
the electric field-autocorrelation function (EACF) of the scattered light
\begin{equation}
\label{eq:EACF-scattered}
    g_E(\qq,\tau) =
    \left\langle \;\! E_s(\qq,\tau) \cdot E_s^\ast(\qq,0)\;\!\right\rangle
    \,.
\end{equation}
The mixed-field IACF contains also a homodyne part, equal to
the homodyne intensity autocorrelation function,
\begin{equation}
\label{eq:IACF-scattered}
    g_I(\qq,\tau) = C_\textrm{hom}(\qq,\tau) =
    \left\langle {|E_s(\qq,\tau)|}^2 \cdot {|E_s(\qq,0)|}^2
    \right\rangle \,
\end{equation}
which is of order ${\cal O}(E_s^4)$. We further note that
$g_E(\qq,0) = \left\langle I_s(\qq) \right\rangle$, and $g_I(\qq,0)
= \left\langle I_s^2(\qq) \right\rangle$. In standard heterodyne
experiments, $\langle I_s \rangle \ll I_r$, and  the homodyne contribution is
negligibly small. Eq. (\ref{eq:IACFmixed-final}) reduces then to
\begin{eqnarray}
\label{eq:IACFmixed-smallEs}
    C_\textrm{shet}(\qq,\tau) = I_r^2 + 2\;\! I_r\;\! \left\langle
    I_s(\qq) \right\rangle \left(\;\! 1 +  \textrm{Re}
    \left[\; \widehat{g}_E(\qq,\tau)\;\!\exp\{- i \Delta\omega_B \tau \}\;\!\right] \;\right)  \,,
\end{eqnarray}
with the normalized EACF of the scattered light,
\begin{eqnarray}
\label{eq:EACF-scattered-normalized}
    \widehat{g}_E(\qq,\tau) = \frac{g_E(\qq,\tau)}{\left\langle
    I_s(\qq) \right\rangle} \,\,,
\end{eqnarray}
defined such that $\widehat{g}_E(\qq,0)=1$. The EACF is
complex-valued for a non-isotropic system in an external
field.

\subsection{Scattered field with Gaussian statistics}
\label{subsec:Gaussian statistics}

So far, we have not assumed that the scattered field is a complex
central Gaussian random variable, an assumption frequently made in
dynamic light scattering theory. This assumption is certainly valid
for dilute suspensions with weak particle correlations,
where it becomes a consequence of the central limit theorem. It is
violated, however, in very dense suspensions of glass-like or
gel-like character that show dynamic heterogeneity
\cite{Furukawa:02}, and in live sperm cells
. For a scattered field of Gaussian character, Eq.
(\ref{eq:IACFmixed-final}) can be derived alternatively using Wick's
theorem for the average of products of central Gaussian random
variables \cite{Dhont book}. Another consequence of Wick's theorem is
that for a Gaussian scattered field, $g_I(\qq,\tau)$ includes no
information on the suspension dynamics, not contained already
in $g_E(\qq,\tau)$, and that they are related by the Siegert relation
\begin{eqnarray}
\label{eq:Siegert}
    \widehat{g}_I(\qq,\tau) = 1 + {| \widehat{g}_E(\qq,\tau) |}^2
    \,\,,
\end{eqnarray}
with $\widehat{g}_I(\qq,0) = 2$.
The mixed-field IACF for a scattered field of Gaussian statistics is
thus
\begin{eqnarray}
\label{eq:IACFmixed-final-Gaussian}
    C_\textrm{shet}(\qq,\tau) &=& \left[ I_r + \left\langle
    I_s(\qq) \right\rangle \right]^2 + 2 I_r \left\langle
    I_s(\qq) \right\rangle \textrm{Re} \left[\;\!\widehat{g}_E(\qq,\tau)
    \exp\{-\;\!i\;\! \Delta\omega_B\;\! \tau \}\;\!\right] \nonumber \\
    & & + \;\!
    {\left\langle I_s(\qq) \right\rangle}^2\;\!{|\widehat{g}_E(\qq,\tau)|}^2
\end{eqnarray}
with the normalized scattered-field IACF,
\begin{eqnarray}
\label{eq:IACF-scattered-normalized}
    \widehat{g}_I(\qq,\tau) = \frac{g_I(\qq,\tau)}{{\left\langle
    I_s(\qq) \right\rangle}^2} \,,
\end{eqnarray}
defined such that $\widehat{g}_I(\qq,0) = \left\langle I_s^2(\qq) \right\rangle /
{\left\langle I_s(\qq) \right\rangle}^2 \geq 1$.
For a non-Gaussian scattered light field, however, one has to deal with
two genuinely different correlation functions
$g_E(\qq,\tau)$ and $g_I(\qq,\tau)$.

\subsection{Super-heterodyne power spectrum}
\label{subsec:power spectrum}

When a spectrum analyzer is used in place of a time correlator to
analyze the superposition of scattered and reference light, the
power spectrum is determined in place of the mixed-field IACF. The
power spectrum is the temporal Fourier transform of the mixed-field
IACF, namely
\begin{eqnarray}
\label{eq:power-spectrum}
    C_\textrm{shet}(\qq,\omega) \;\!=\;\! \frac{1}{2\pi}
    \int_{-\infty}^{\infty} \!d\tau\;\! \exp\{i\;\!\omega\;\!\tau
    \}\;\! C_\textrm{shet}(\qq,\tau)  \;\!=\;\! \frac{1}{\pi}
    \int_0^{\infty} \!d\tau\;\! \cos\{\omega\;\!\tau\}\;\! C_\textrm{shet}(\qq,\tau) \,,
\end{eqnarray}
where the second equality follows from the stationarity property
$C_\textrm{shet}(\qq,\tau) =  C_\textrm{shet}(\qq,-\tau)$. Stationarity implies further
that $g_E(\qq,\tau) = g_E(\qq,-\tau)^\ast$ and $g_I(\qq,\tau) = g_I(\qq,-\tau)$.
The mixed-field IACF and its power spectrum are measurable
quantities and therefore real-valued. The power spectrum is symmetric in $\omega$, that is
it consists of two sub-spectra extending symmetrically to positive and negative frequencies. Furthermore,
it fulfills the sum rule
\begin{eqnarray}
\label{eq:sum-rule}
    \int_{-\infty}^{\infty} \!d\omega\;\! C_\textrm{shet}(\qq,\omega) \;\!=\;\! C_\textrm{shet}(\qq,\tau = 0) = (I_r)^2 + 4\;\!I_r \langle I_s(\qq) \rangle + \langle I_s(\qq)^2 \rangle \,.
\end{eqnarray}
The 'static' quantity $\langle I_s(\qq) \rangle = \langle I_s(-\qq) \rangle$ in a driven, non-equilibrium system
still depends on the hydrodynamic interactions \cite{Leshansky:08}.

\subsection{Non-interacting colloidal spheres}
\label{subsec:non-interacting}

Consider a very dilute charge-stabilized suspension, which includes a
sufficient amount of added salt ions, so that the colloidal
spheres are non-interacting. Suppose that
the $N$ spheres in the scattering volume form $m$
components, with $N_\alpha$ particles in component $\alpha
\in \{1,...,m\}$. The position vectors of
non-interacting spheres are independent so that
\begin{eqnarray}
   \left\langle I_s^{\;o}(\qq) \right\rangle \;\!=\;\! {|c|}^2
   \sum_{\alpha=1}^m \;\!N_\alpha \;\!f_\alpha^2(q)  \,,
\end{eqnarray}
and
\begin{eqnarray}
   \widehat{g}_E^{\;o}(\qq,\tau) \;\!=\;\! \frac{
   \sum_{\alpha=1}^m \;\!x_\alpha \;\!f_\alpha^2(q)\!\; G_\alpha^0(\qq,\tau) }
   {\sum_{\alpha=1}^m \;\!x_\alpha \;\!f_\alpha^2(q) }  \,.
\end{eqnarray}
Here, $x_\alpha = N_ \alpha/N$ is the molar fraction, $f_\alpha(q)$ the scattering amplitude, and
\begin{eqnarray}
\label{eq:self-factor-zero}
   G_\alpha^{\;o}(\qq,\tau) \;\!=\;\! \left\langle \exp\{i\;\!{\qq} \cdot ({\bf r}_\alpha(\tau)
   -{\bf r}_\alpha(0) \} \right\rangle \,,
\end{eqnarray}
the self-intermediate scattering function of component
$\alpha$. The label $(^o)$ indicates properties
of a system of non-interacting particles.

The self-intermediate scattering function is the spatial Fourier
transform of the fundamental solution $P_s(\rr,\tau|\rr_0)$, with
$P_s(\rr,0|\rr_0)= \delta(\rr - \rr_0)$, of the Smoluchowski equation,
\begin{eqnarray}
\label{eq:GSE-single}
   \frac{\partial}{\partial \tau} P_s(\rr,\tau) \;\!=\;\! D_0 \nabla^2 P_s(\rr,\tau)
    - \vv_e^{o} \cdot \nabla P_s(\rr,\tau) \,,
\end{eqnarray}
which is used to describe a single microion-dressed charged colloidal sphere
of diffusion coefficient $D_0$ in an unbounded and quiescent fluid subject
to a constant electric field ${\bf E}$.
The colloidal macroion responds to this forcing by acquiring a terminal
mean drift velocity relative to the quiescent solvent
\begin{eqnarray}
\label{eq:mobility-single}
   \vv^{o}_e \;\!=\;\! \mu_e^{o} \;\!{\bf E}\,,
\end{eqnarray}
once the polarization of the microionic electric double layers (EDL)
in response to the external field has acquired a steady state
after the relaxation time
$\tau_\textrm{DL} = (\kappa^2\;\!\!D_\textrm{el})^{-1}$, which is
typically $10^{-7}$ s for a 1 mM electrolyte solution \cite{Hunter:Colloid_II_book}.
Here, $\mu_e^o$ is the electrophoretic mobility of a colloidal macroion
in isolation, and $\kappa^{-1}$ and $D_\mathrm{el}$ are the Debye
screening length and mean diffusion coefficient, respectively, of
electrolyte ions. The assumption underlying the use of the forced diffusion Eq. (\ref{eq:GSE-single})
is that the particle diffusion and the electrokinetic effects on the EDL caused by the applied field
are independent. This assumption is reasonable for a weak field which can be treated as a first-order
perturbation (linear electrophoresis). The field-independent mobility of a spherical colloid
can be calculated using
standard electrokinetic mean-field theory (see, e.g., \cite{Hunter:Colloid_II_book,Ohshima:book}) or
extensions which account for microion correlations \cite{LozadaCassou:01}. These theories provide a
relation between the experimentally determined electrophoretic mobility and the zeta potential or related
electrophoretic colloid charge.

The fundamental solution of the single-colloid Smoluchowski equation is
\begin{eqnarray}
\label{eq:fundamental}
   P_s(\rr,\tau \geq 0|\rr_0) \;\!=\;\! \frac{1}{\left(4 \pi D_0
   \tau\right)^{3/2}} \;\!
   \exp\Big\{ -\frac{\left[\rr -\rr_0 -\vv_e^{o} \tau \right]^2}{4 D_0 \tau} \Big\}
   \,,
\end{eqnarray}
stating that $\rr -\rr_0 -\vv_e^{o} \tau$ is a central
Gaussian random variable so that $g_I^{\;o}(\qq,\tau)$ is related
to $g_E^{\;o}(\qq,\tau)$ by the Siegert relation.

Eq. (\ref{eq:fundamental}) applies also to a polydisperse system of
statistically independent spheres of varying electrophoretic drift velocities.
Electrophoresis differs from
sedimentation in that the total force exerted on the
suspension is zero due to overall electro-neutrality. This has
consequences for the range of the HI, which is shorter in the case of
electrophoresis \cite{Allison:00,Long_Ajdari_screening:01}.

The self-intermediate scattering function of an isolated sphere of
diffusion coefficient $D_0 = k_BT/(6\pi \eta_0 a)$, which drifts
with the mean velocity $\vv_e^{o} = \mu_e^o \;\!{\bf E}$, follows
from Fourier-transforming the fundamental solution,
\begin{eqnarray}
   G^{\;o}(\qq,\tau \geq 0) &=&
   \int\;\!\!  d\!\left(\rr - \rr_0\right) \;\! \exp\{i \;\! \qq \cdot \left(\rr -\rr_0 \right)\} P_s(\rr,\tau|\rr_0)
   \nonumber \\
     &=&   \exp\{i\;\!\qq\cdot\vv_e^{o}\;\! \tau\}
   \exp\{ - q^2\;\!D_0\!\;\tau\}
    \,,
\end{eqnarray}
where the sign convention in the Fourier transform has been made consistent with the one in Eq. (\ref{eq:self-factor-zero}).

In summary, the normalized EACF and IACF of a monodisperse
suspension of non-interacting colloidal spheres with equal
electrophoretic drift velocity $\vv_e^{o}$ is given, for all $\tau$,
by
\begin{eqnarray}
   \widehat{g}_E^{\;o}(\qq,\tau) &=&
   \exp\{i\;\!\qq\cdot\vv_e^{o}\;\! \tau
   \}\;\! \exp\{ - q^2\;\!D_0\!\; |\tau| \} \\
   \widehat{g}_I^{\;o}(\qq,\tau) &=&
   1 + \exp\{ - 2\;\!q^2\;\!D_0\!\; |\tau| \} \,,
\end{eqnarray}
which shows that the homodyne part of the mixed-field IACF is
independent of $\vv_e^{o}$ as long as no differential motion between
two spheres due to different drift velocities occurs. Polydispersity
in the particle mobility would introduce such a differential motion
between different components $\alpha$ and $\beta$ and thus be detectable as an
oscillation in $\widehat{g}_I^{\;o}(\qq,\tau)$ \cite{Robertson:91}. Continuing with the
case of equal drift velocities and using Eq.
(\ref{eq:IACFmixed-final-Gaussian}), the mixed-field IACF becomes
\begin{eqnarray}
\label{eq:IACFmixed-final-Gaussian-zero}
    C_\textrm{shet}^{\;o}(\qq,\tau) \!&=& \left[ I_r + \left\langle
    I_s^o(\qq) \right\rangle \right]^2 + 2\;\! I_r \left\langle
    I_s^{\;o}(\qq) \right\rangle \;\!\cos{\left[\;\!
     \left( \Delta\omega_B - \qq\cdot\vv_e^{o} \;\! \right) \tau
     \;\!\right]} \;\! \exp\{ -
     q^2\;\!D_0\!\; |\tau| \}
     \nonumber \\
     & & + \;\!{\left\langle I_s^{\;o}(\qq) \right\rangle}^2 \;\! \exp\{ -
     2\;\!q^2\;\!D_0\!\; |\tau| \}\,,
\end{eqnarray}
where $I_s^{\;o}(\qq) = |c|^2\;\!N\;\!f^2(q)$. Suppose now that $\left\langle I_s^{\;o} \right\rangle
\ll I_r$. Then,
\begin{eqnarray}
\label{eq:IACFmixed-final-Gaussian-zero-smallIs}
    C_\textrm{shet}^{\;o}(\qq,\tau) \;\!=&\;\!  I_r^2 + 2\;\!I_r\;\! \left\langle
    I_s^o(\qq) \right\rangle + 2 \;\! I_r \left\langle
    I_s^{\;o}(\qq) \right\rangle \;\! \cos{\left[\;\!
     \left( \Delta\omega_B - \qq\cdot\vv_e^{o} \;\! \right) \tau
     \;\!\right]} \;\! \exp\{ -
     q^2\;\!D_0\!\; |\tau| \} \,,
\end{eqnarray}
i.e., provided the homodyne part is negligibly,
$C_\textrm{shet}^{\;o}(\qq,\tau)$ is an exponentially damped cosine
in $\tau$, shifted upward by a time-independent baseline, and
characterized by a
\begin{eqnarray}
    \textrm{decay time:} && \tau_q = \frac{1}{q^2\;\!D_0} \nonumber \\
     \textrm{period:} && T = \frac{2\;\!\pi}{|\Delta\omega_B - \qq\cdot\vv_e^{o}| } \,.
\end{eqnarray}
Thus, a measurement of the mixed-field IACF
provides a simultaneous determination of the
diffusion coefficient and the electrophoretic mobility, given in
our setup by $\mu^o_e = |(\qq \cdot \vv_e^{o})|/(q\;\!
E)$. Super-heterodyning using Bragg cells has two effects not
present in a  conventional heterodyne setup:
First, through the appearance of a non-zero $\Delta\omega_B$, the oscillation period $T$ can be
shortened relative to the decay time
$\tau_q$, which allows for an improved resolution in $v_e^0$, and
second, the method is sensitive to
the sign of $\vv_e^{o}$. A drift velocity parallel to $\qq
\parallel {\bf E} \parallel \widehat{\bf z}$ or, more generally, a velocity $\vv_e^0$ sharing an acute angle
with $\qq$, enlarges the period of oscillations,
whereas antiparallel motion causes the period to decrease. In our setup, the negatively
charged colloids move antiparallel to the $\qq$-axis (i.e., $- \qq \cdot \vv_e^0 = q\;\!|v_e^0| > 0$),
so that a reduction in the period $T$ is
achieved. The resolution can be further improved by making $\tau_q$ larger, that is by
choosing the scattering angle as small as possible. For typical
colloids, $\tau_q^{-1} = \Delta\omega \sim 10^{3}$ Hz, so that the symmetric
frequency broadening due to the isotropic diffusive motion of spheres is
comparable to the frequency shift by the Bragg cells.

Using Eq. (\ref{eq:power-spectrum}), the mixed-field power spectrum is obtained as
\begin{eqnarray}
\label{eq:powerspectrum-zero}
    C_\textrm{shet}^{\;o}(\qq,\omega) && \!\! = \left[ \;\! I_r + \left\langle
    I_s^o(\qq) \right\rangle \;\! \right]^2 \;\! \delta(\omega) \nonumber \\
    && + \; \frac{ I_r \left\langle
    I_s^{\;o}(\qq) \right\rangle }{\pi}
    \left[ \frac{q^2\;\!D_0}{ \left(\omega + \left[\Delta\omega_B - \qq \cdot \vv_e^{o}\right] \right)^2 +
    \left( q^2\;\!D_0\right)^2 }  +
    \; \frac{q^2\;\!D_0}{ \left(\omega - \left[\Delta\omega_B - \qq \cdot \vv_e^{o}\right] \right)^2 +
    \left( q^2\;\!D_0 \right)^2 } \right] \nonumber \\
    && + \; \frac{{\left\langle
    I_s^{\;o}(\qq) \right\rangle}^2}{\pi} \frac{2 \;\! q^2 \;\! D_0}{ \omega^2 + \left( 2 \;\! q^2 \;\! D_0
    \right)^2 }  \,.
\end{eqnarray}

It is the sum of an irrelevant singular term at $\omega=0$ due to the
non-subtracted baseline contribution, and two symmetrically shifted
Lorentzian curves of half-width at half-height $q^2\;\!D_0$,
centered at  $\omega = \pm \left[\Delta\omega_B - \qq \cdot
\vv_e^{o} \right]$, which is the interesting super-heterodyne part,
and an unshifted Lorentzian of double-sized half-width at
half-height $2\;\!\! q^2\;\!\!D_0$, centered at $\omega=0$. The last
contribution is due to the self-beating of the scattered light (homodyne
part) and contains no information on the electrophoretic motion.
The two heterodyne Lorentzians are Bragg-shifted away from the
origin by $\Delta\omega_B$, so that they can be better distinguished
from the central homodyne Lorentzian. Since ${\bf q}$ is parallel to
${\bf E} = E \;\!\widehat{\bf z}$ in our setup, and since the negatively charged
colloids move antiparallel to the field, the centers of the super-heterodyne Lorentzians are
separated from the frequency origin by $\Delta\omega_B + q\;\!\mu_e^0\;\! E$,
so that this separation increases with increasing field strength.
The frequency resolution improves with decreasing $q$ and $D_0$, and
increasing field strength $E$. Note that the frequency shift in the mixed-field power
spectrum corresponds to the sinusoidal modulation in the mixed-field EACF.

When the electrophoretic mobilities vary from particle to particle possibly due to some charge
polydispersity, the super-heterodyne part of the mixed-field IACF in Eq. (\ref{eq:IACFmixed-final}) is generalized to
\begin{eqnarray}
\label{eq:mobility-polydispersity1}
   C_\textrm{shet}^{\;o}(\qq,\tau)|_\textrm{part} \;\!=\;\! 2\;\! I_r \left\langle
    I_s^{\;o}(\qq) \right\rangle \textrm{Re} \left[\;\!   {\langle\;\widehat{g}_E^o(\qq,\tau) \;\rangle}_{\mu_e}
    \exp\{-\;\!i\;\! \Delta\omega_B\;\! \tau \}\;\!\right] \,,
\end{eqnarray}
where $\langle \cdots \rangle_{\mu_e}$ denotes an average over the mobility distribution. A narrow distribution characterized by
the mean mobility $\overline{\mu}_e$ and the relative standard deviation $\sigma_{\mu_e}$, can be approximated by a Gaussian, yielding \cite{Robertson:91}
\begin{eqnarray}
\label{eq:mobility-polydispersity2}
   C_\textrm{shet}^{\;o}(\qq,\tau) |_\textrm{part} = 2\;\! I_r \left\langle
    I_s^{\;o}(\qq) \right\rangle \exp\{ - q^2 \left[ D_0\!\;|\tau| + \frac{1}{2} \left(\overline{\mu}_e E \tau \right)^2 \sigma^2_{\mu_e} \right] \} \cos{\left[\;\!
     \left( \Delta\omega_B + q\overline{\mu}_e E \;\! \right) \tau
     \;\!\right]}  \,,
\end{eqnarray}
for non-interacting colloidal spheres. Hence the super-heterodyne part of the IACF decays faster than exponentially with a rate
that increases with increasing mobility polydispersity $\sigma_{\mu}$. The corresponding power spectrum has a shape similar to a
Lorentzian centered at $\Delta\omega_B + q\overline{\mu}_e E$, but is broadened symmetrically by an amount that increases with
increasing $\sigma_{\mu_e}$.

A remark on the sign convention for $\qq$ made in Eq. (\ref{eq:scattering-vector}) is in order here. In fact, we
could have equally well used the definition $\qq = -({\bf k}_i - {\bf k}_s)$ for the scattering wave vector, which
amounts to replacing $\qq$ by $-\qq$ in Eq. (\ref{eq:Escattered}). The
only consequence is that $g_E(\qq,\tau)$ must be replaced by $g_E(-\qq,\tau) = g_E(\qq,\tau)^\ast$
in all the following expressions, on noting further that $g_I(-\qq,\tau) = g_I(\qq,\tau)$ and
$\langle I_s(-\qq) \rangle = \langle I_s(\qq) \rangle$. For non-interacting particles, this leads to the
replacement of $\Delta\omega_B - \qq \cdot \vv_e^0$ by $\Delta\omega_B + \qq \cdot \vv_e^0$ in
Eqs. (\ref{eq:IACFmixed-final-Gaussian-zero} - \ref{eq:powerspectrum-zero}).
This makes it obvious that the outcome of the experiment is independent on the sign convention used for the scattering vector.

\subsection{Interacting colloidal spheres}
\label{subsec:interacting}

Information on the colloid velocities is embodied only in the super-heterodyne part of $C_\textrm{shet}(\qq,\tau)$
that involves $g_E(\qq,\tau)$.
The normalized EACF of light scattered from $N$ interacting
spheres is given by
\begin{eqnarray}
\label{eq:EACF-manyspheres}
   \widehat{g}_E(\qq,\tau) \;\!=\;\!
    \frac{\sum_{l,j=1}^N f_l(q)\;\!f_j(q)\;\!
    \left\langle\;\! \exp\{i\;\!\qq \cdot \left[\rr_l(\tau) - \rr_j(0) \right] \} \;\!\right\rangle}
    {\sum_{l,j=1}^ N f_l(q)\;\!f_j(q)\;\!
    \left\langle\;\! \exp\{i\;\!\qq \cdot \left[\rr_l - \rr_j \right] \} \;\!\right\rangle}
    \,.
\end{eqnarray}
Consider first an ideally monodisperse suspension of identical
spheres. Then,
\begin{eqnarray}
   \widehat{g}_E(\qq,\tau) \;\!=\;\! \frac{S(\qq,\tau)}{S(\qq)}
   \,,
\end{eqnarray}
and
\begin{eqnarray}
\label{eq:Isvonq}
   \left\langle\;\! I_s(\qq)\,\! \right\rangle  \;\!=\;\! {|c|}^2\;\!N\;\! f^2(q)\;\! S(\qq)   \,.
\end{eqnarray}
Here,
\begin{eqnarray}
   S(\qq,\tau) \;\!=\;\! \left\langle\;\! \frac{1}{N} \sum_{l,j=1}^N \;\!
    \exp\{i\;\!\qq \cdot \left[\rr_l(\tau) - \rr_j(0) \right] \} \;\! \right\rangle
\end{eqnarray}
is the complex-valued, steady-state dynamic structure factor.
The corresponding steady-state structure factor is $S(\qq) =
S(\qq,0)$.

For a $m$ component mixture, it follows from Eq. (\ref{eq:EACF-manyspheres}) that

\begin{eqnarray}
\label{eq:EACF-manycomponents}
   \widehat{g}_E(\qq,\tau) \;\!=\;\!
    \frac{ \sum_{\alpha,\beta=1}^m \left(x_\alpha\;\!x_\beta\right)^{1/2} \;\! f_\alpha(q)\;\!f_\beta(q)\;\!
    S_{\alpha \beta}(\qq,\tau) }
    { \sum_{\alpha,\beta=1}^m \left(x_\alpha\;\!x_\beta\right)^{1/2} \;\! f_\alpha(q)\;\!f_\beta(q)\;\!
    S_{\alpha \beta}(\qq) }
    \,,
\end{eqnarray}
with the partial dynamic structure factor of $\alpha-\beta$
colloid pairs defined by
\begin{eqnarray}
   S_{\alpha\beta}(\qq,\tau) \;\!=\;\! \left\langle\;\! \frac{1}{\left(N_\alpha N_\beta \right)^{1/2}}
    \sum_{l=1}^{N_\alpha} \sum_{j=1}^{N_\beta}\;\!
    \exp\Big\{ i\;\!\qq \cdot \left[\rr_l^\alpha(\tau) - \rr_j^\beta(0) \right] \Big\} \;\!
    \right\rangle \,,
\end{eqnarray}
where $\rr_l^\alpha$ points to the center of a sphere $l$
belonging to component $\alpha$. The partial dynamic
structure factor can be written as the sum of a self- and
distinct part,
\begin{eqnarray}
\label{self-distinct}
   S_{\alpha\beta}(\qq,\tau) \;\!=\;\! \delta_{\alpha\beta}\;\! G_\alpha(\qq,\tau) + S^d_{\alpha\beta}(\qq,\tau)
    \,,
\end{eqnarray}
where the self-intermediate scattering function of an $\alpha$-type sphere is
defined such that $G_\alpha(\qq,0) = 1$.
On invoking the definition
\begin{eqnarray}
\label{Smeasured}
   S_M(\qq,\tau) \;\!=\;\! \frac{1}{\overline{f^2(q)}}
   \sum_{\alpha,\beta = 1}^m \left( x_\alpha x_\beta \right)^{1/2}\;\! f_\alpha(q)\;\! f_\beta(q)\;\! S_{\alpha\beta}
   (\qq,\tau)
\end{eqnarray}
of the measurable dynamic structure factor in a polydisperse system, and
\begin{eqnarray}
\label{fquadrat}
   \overline{f^p}(q) \;\!=\;\! \sum_{\alpha = 1}^m x_\alpha {f_\alpha(q)}^p \,,
\end{eqnarray}
for any integer $p$, Eq. (\ref{eq:EACF-manycomponents}) is rewritten in more compact notation as
\begin{eqnarray}
\label{eq: EACF-manycomponents-compact}
   \widehat{g}_E(\qq,\tau) \;\!=\;\!
    \frac{S_M(\qq,\tau)}{S_M(\qq)}
    \,,
\end{eqnarray}
with $S_M(\qq)=S_M(\qq,0)$.

Suppose now that the $m$ components differ only in their scattering
amplitudes $\{f_\alpha(q)\}$, but are otherwise identical. For
such an interaction-monodisperse system, the partial dynamic
structure factors reduce to \cite{Nagele:96}
\begin{eqnarray}
   S_{\alpha\beta}(\qq,\tau) \;\! = \;\! \delta_{\alpha\beta} \;\! G(\qq,\tau) +
   \left( x_\alpha \;\! x_\beta \right)^{1/2} \;\!
   \left[ S(\qq,\tau) - G(\qq,\tau) \right] \,,
\end{eqnarray}
where $G(\qq,\tau)$ and $S(\qq,\tau)$ are, respectively, the
self-intermediate scattering function and the dynamic structure
factor of an ideally monodisperse system. Therefore, for pure scattering polydispersity,
the EACF of the scattered light reduces to
\begin{eqnarray}
    g_E(\qq,\tau) \;\!=\;\! {|c|}^2\;\!N\;\! \overline{f^2}(q) \;\!S_D(\qq,\tau) \,,
\end{eqnarray}
with the so-called decoupling (approximation) dynamic structure factor
\begin{eqnarray}
\label{eq:decoupling-Svonq}
   S_D(\qq,\tau) \;\!=\;\! X(q) \;\! G(\qq,\tau) + \left[1 - X(q) \right]\;\! S(\qq,\tau) \,,
\end{eqnarray}
as an approximation to $S_M({\bf q},\tau)$, and the decoupling scattering factor
\begin{equation}
   X(q) \;\! = \;\! 1 - \frac{ \overline{f(q)}^{\;2} }{ \overline{f^2(q)} } \,.
\end{equation}
For a suspension  of slightly size-polydisperse homogeneous spheres made of the
same scattering material, $X(q)$ simplifies to
\begin{eqnarray}
\label{X-simplified}
    X(q) \;\! \approx \;\! 9 \;\! s^2 \,,
\end{eqnarray}
provided that $s < 0.1$ and $q\;\!\! \sigma < 1.5$, where $\sigma = 2 a$ is the
mean diameter.
The polystyrene sphere samples explored in this work are only slightly size-polydisperse,
characterized  by the relative standard deviation of $s \approx 0.05$.
For these systems, the assumption of interaction-monodispersity is thus quite reasonable, and it implies
for $9\;\!s^2 \ll 1$, and in the experimentally probed small-$q$ range, that
\begin{eqnarray}
\label{eq:self-part-scattering}
    g_E(\qq,\tau) \;\! \approx \;\!   {|c|}^2 \;\! N\;\! f^2(q) \left [ 9 \;\! s^2 \;\! G(\qq,\tau) + S(\qq,\tau) \right] \,,
\end{eqnarray}
where terms of $O(s^4)$ and $O(s^2\times S(\qq))$ have been ignored. Here, $f(q)$ is the scattering amplitude of
ideally monodisperse spheres of diameter $\sigma$.

For values of $q$ well below the position, $q_m$,
of the principal peak, the zero-field $S(q)$ of deionized suspensions
is very small in comparison to one, since the osmotic compressibility is very
low. In these deionized fluid systems, $q_m \approx 2\pi n^{1/3} = \left(6 \pi^2
\phi \right)^{1/3}/a$, where $\phi$ is the volume fraction of spheres \cite{banchio das neueste}.

On assuming that $S(\qq)$ remains very small for $|\qq| \ll q_m$,
also in the presence of a weak external field, as it can be expected in the linear response regime,
we obtain approximately
\begin{eqnarray}
\label{eq:self-part-only}
    \widehat{g}_E(\qq,\tau) \;\!\approx\;\! G(\qq,\tau) \,,
\end{eqnarray}
provided that $q \ll q_m$ and $S(\qq) \ll 9\;\!s^2$. The scattered-field EACF in the
super-heterodyne part of Eq. (\ref{eq:IACFmixed-final}) is then determined
essentially by the self-intermediate scattering function, without
significant contributions from $S(\qq,\tau)$.

We point out here that the field-free coherent scattering
contribution, $S(q,\tau)$, to $S_D(q,\tau)$ decays faster than
$G(q,\tau)$ at low $q$, even at longer times
when the slowing influence of
particle caging comes into play. This is due to the large value of the (field-free) collective
diffusion coefficient, $D_c = D_0\;\!H(q \to 0)/S(q\to 0)$ in comparison to $D_0$, where
$H(q)$ is the so-called hydrodynamic function \cite{banchio das neueste}.
In fact, $G(q \ll q_m,\tau) \approx \exp\{ -q^2 \;\! W(\tau) \}$ decays
slowly at small $q$ since
the particle mean-squared displacement, $W(\tau)$, has a long-time slope equal to the long-time
self-diffusion coefficient, that is significantly smaller than $D_0$. For an example,
near the freezing transition, the long-time self-diffusion coefficient amounts to only about
ten percent of $D_0$.
Therefore,  $G(\qq,\tau)$ is expected to dominate
$S(\qq,\tau)$ even in presence of a weak external field.

It is a difficult task to calculate the dynamic structure factor,
$S(\qq,\tau)$, of interacting macroions in the presence
of an electric field since this quantity depends on all ionic
degrees of freedom. In fact, to our knowledge, no general theoretical or simulation results
are available to date for the dynamic structure factor of interacting charged colloids
in an external electric field. Results are known only for very simplifying models of driven colloid systems.
To give an example, the stationary dynamic structure factor of a homogeneous and ideally
monodisperse suspension of slowly sedimenting
colloidal spheres where (quite unrealistically) hydrodynamic interactions have been ignored
is given by $S(\qq,\tau) = S_\textrm{eq}(q,\tau)
\times \exp\{i\;\!{\bf q} \cdot {\bf V}_\textrm{sed}\;\!\tau \}$. Here, ${\bf V}_\textrm{sed}$ is the sedimentation velocity
measured in the laboratory frame,
which is corrected for the solvent backflow originating from the presence of the container bottom, and $S_\textrm{eq}(q
,\tau)$ is the equilibrium dynamic structure factor of the non-driven system. Many-body hydrodynamic interactions
and polydispersity in a suspension of particles subject to a constant forcing cause a change of the equilibrium microstructure
into a new non-equilibrium state so that the use of a simple Galilei transformation becomes invalid.
In electrophoretic experiments, the electro-kinetic coupling of ions and possibly existing electro-osmotic solvent flow
are additional complications that add to the complexity in calculating $S(\qq,\tau)$.

To rationalize heuristically how the electrophoretic mobility of strongly interacting
particles may be determined from a low-$q$ measurement of $G(\qq,\tau)$,
we make the assumption, just like for non-interacting particles, that
diffusion and electrophoretic drift are uncoupled in a weak external field.
We impose this assumption on a coarse-grained level only, where
distances larger than $\Delta x \gg 2\pi/q_m$ are resolved, corresponding to long
correlation times $\tau \gg \tau_m=(q_m^2\;\!D_0)^{-1}$.
On ignoring in addition polydispersity effects, the coarse-grained single-particle mean current, $\overline{j}_s(\rr,\tau)$,
associated with coarse-grained single-particle density $\overline{\rho}_s(\rr,\tau)$, can be decomposed as
\begin{eqnarray}
\label{self-flux}
    \overline{j}_s(\rr,\tau) \;\!=\;\! - D_s(\phi)\;\! \nabla
    \overline{\rho}_s(\rr,\tau) +
    \vv_e(\phi)\;\overline{\rho}_s(\rr,\tau) \,.
\end{eqnarray}
In this phenomenological description, $D_s(\phi)$ is the unperturbed long-time self-diffusion coefficient, and
$\vv_e(\phi)$ is identified with the long-time mean electrophoretic
velocity of the spheres. In conjunction with the continuity equation,
\begin{eqnarray}
\label{self-continuity}
    \frac{\partial}{\partial\;\!\tau} \;\! \overline{\rho}_s(\rr,\tau) +
    \nabla \cdot \overline{j}_s(\rr,\tau) \;\!\ = \;\! 0 \,,
\end{eqnarray}
the single-particle eq. (\ref{eq:GSE-single}) is recovered, but $D_s(\phi)$ is replacing now $D_0$, and ${\bf v}_e(\phi)$
replaces the single-sphere electrophoretic velocity ${\bf v}_e^0$.
The self-intermediate scattering
function is here given by
\begin{eqnarray}
\label{eq:Markovian}
   G(|\qq| \ll q_m,\tau \gg \tau_m) = \langle
   \overline{\rho}_s(\qq,\tau) \;\! \overline{\rho}_s(-\qq,0)
   \rangle \approx
   \exp\{i\;\!\qq\cdot\vv_e(\phi)\;\! \tau
   \}\;\!
   \exp\{ - q^2\;\!D_s(\phi)\!\;\tau\}
    \,.
\end{eqnarray}
For deionized
aqueous suspensions of PnBAPS68 spheres the system parameters are $\sigma = 68$ nm,
$\phi \approx 10^{-3}$, $T = 293.15$ K, $\nu_s(\lambda,T = 20 C) \approx 1.33$ (water),
$\lambda = 488$ nm in the static structure
factor measurements for zero external field,
and $\lambda = 532$ nm in the electrophoresis setup with $|{\bf E}| \approx 10 - 120$ V/cm.
This gives  $\tau_a = a^2/D_0
= 0.18$ ms, $D_0 = 6.3 \;\! \mu\mathrm{m}^{2}\mathrm{/s}$, $\left(q^\ast\right)^2 D_0 = 156$ Hz corresponding
to $\tau_{q^\ast} = 6.4$ ms, and $q^\ast/q_m \approx 0.4$ where
$q^{\ast} = ( 4\pi\nu_s /\lambda ) \sin\{ \Theta/2\} = 4.97 \;\!\mu\mathrm{m}^{-1}$ is the wave
number used in the electrophoretic setup. The correlation times are in the range of
$10 - 100$ ms or larger so that long-time particle motion is probed.

The assumption on the decoupling of diffusion and drift
is flawed when the
electric field is so strong that $D_s(\phi)$ and ${\bf v}_e(\phi)$
become field-dependent \cite{Barany:04}.
To test the validity of Eq. (\ref{eq:Markovian}) in a cell geometry
with negligible osmotic flow requires to verify whether
the super-heterodyne peaks in
$C_\textrm{shet}(\qq,\omega)$ are of a Lorentzian shape and, if this
has been confirmed, to determine the diffusion coefficient from the
measured half-width at half-height, $\Delta \omega_{1/2}$,
according to $D_s = \Delta\omega_{1/2}/q^2$. This
coefficient should be substantially smaller than $D_0$ in the case
of a deionized system, provided it can be identified with the unperturbed
long-time self-diffusion coefficient. The short-time
self-diffusion coefficient in these systems, on the other hand, is only slightly
smaller than $D_0$, with the concentration dependence given by
$D_0 \left( 1 - a_t\;\! \phi^{4/3} \right)$, where $a_t \approx 2-3$ \cite{banchio das neueste}.

Finally, to check measured velocities, the electrophoretic mobility
is determined by the center of the heterodyne Lorentzian
contribution to the power spectrum. It
can be compared to theoretical predictions obtained from different
methods. Quite recently, substantial progress has been made in
developing novel numerical schemes that allow to calculate the
electrophoretic mobility of charged colloids at non-zero
concentrations. These schemes are based, respectively, on a smoothed
surface method, where the sharp colloid-fluid interface is replaced
by a diffuse interface \cite{Kim:06}, on a fluid particle dynamics
(FPD) method with charge-densities included \cite{Tanaka:08}, and a
hybrid simulation method that combines the Lattice Boltzmann fluid
description with a Langevin equation treatment of the colloidal
particles \cite{ELS-PRL,Lobaskin:04}.  A semi-analytic method to
compute the electrophoretic mobility, $\mu_e(\phi)$, in dense systems
based on a simplified mode-coupling scheme applied to the Primitive
Model, has been developed very recently by one of the present
authors \cite{McPhieinprogress}. All these methods allow, in
principle, to explore the electrophoresis of charged colloids with
strongly overlapping electric double layers. Also in analytical
theory, considerable progress was reported in extending the standard
electrokinetic model of dilute systems, and dense systems with
weakly overlapping double layers \cite{Ohshima:book,Levine:74} based
on the single-macroion cell model, to more dense systems of
particles with strong electrostatic interactions \cite{Carrique}.

\subsection{Solvent flow profile}
\label{solvent-flow-profile}

The scattering theory described so far assumes a uniform motion of all spheres. This
is experimentally realized in cells where the electrodes have
a parallel plate capacitor geometry and are placed far away from any
container wall to avoid electro-osmotic solvent flow \cite{Uzgiris}.
Cations in solution tend to concentrate at the negatively charged (glass)
walls and move towards the cathode, dragging solvent (water) with
them. This flow establishes a hydrostatic pressure which tends to
force the center of the fluid to flow towards the anode,
establishing thus a parabolic back flow in the cell \cite{Hunter Zeta}.
It is an open question whether the osmotic flow
profile is affected significantly by a non-zero concentration of strongly
interacting colloidal macro-ions (usually negatively charged) which move towards the anode
and drag also solvent along with them. In a first approximation, however,
this feedback of the colloids on the fluid flow may be ignored. Moreover, as discussed in
\cite{Minor:97}, to built up an electro-osmotic profile
takes quite  a large time of the order $0.1 - 1$ s,
and this time should be accounted for
in experiments with an alternating electric field. In our setup, the switching period is
$15$ s or larger so that the mobility is measured in the steady-state with
a fully developed flow profile.

The electro-osmotic flow profile in the $x-z$ midplane at $y=0$ of a rectangular vessel,
for the solvent streaming along the $z$-axis with the velocity $u_S(x = \pm d) = u_{eo}$
very near the cell walls, has been calculated by
Komagata \cite{Komagata}. On assuming that this profile is macroscopically unperturbed by the interacting colloids,
and that the electrophoretic particle velocity, $v_{e} < 0$, measured relative to the solvent remains constant under the shearing
solvent motion, we obtain using Eq. (\ref{particle-velocity-profile}),
\begin{eqnarray}
\label{eq:Komagata}
    v(x) \;\!=\;\! v_e + u_S(x) \;\!=\;\! v_{e} + u_{eo}\left[ 1 - 3 \;\! \left( \frac{1 -  x^{2}/d^2}
    {2 - 384/(\pi^{5} K)} \right) \right] \,,
\end{eqnarray}
for the velocity profile of the particles in the laboratory (cell)
frame. Here, $K = h/d$, and $\pm d$ and $\pm h$ are the locations of
the cell walls in the $x$ and $y$-direction, respectively. We see
here that the solvent profile $u_S(x)$ is proportional to the
electro-osmotic velocity, $u_{eo} > 0$, which for small fields grows
linearly in $E$. It is thus characterized by the zero-field
mobility, $\mu_{eo} = u_{eo}/|{\bf E}|$, of microionic cations in
the mid-plane close to the cell walls. Again we have assumed equal
mobilities and thus equal $v_e$ for all particles. A distribution of
velocities can conveniently be introduced at this place, when a
coupling of differential velocities and structure is neglected.

In our experiments the aspect  $K$ is equal to $10$. For the
theoretical interpretation of the experimental power spectra of
interacting colloids, we assume now that the homogeneous
distribution of particles remains unperturbed by the locally
shearing motion of the solvent flow profile so that the particle
velocity distribution derives from the Komagata profile according to
$P(v)\;\!dv \propto n \;\! dx$, with $n$ constant. This implies, in
turn, that the solvent profile is macroscopically unperturbed by the
interacting colloids. The explicit form of the normalized $P(v)$
derived from the Komagata profile is given in \cite{TPHV} and will
not be repeated here. In fact, for a sheared suspension of (neutral)
rigid colloidal spheres, Brownian motion will give rise to a uniform
particle distribution. A flow-induced cross-stream migration is
observed, however, for flexible particles like polymers and
polyelectrolytes. At large Peclet numbers, this migration can lead
to a non-uniform center-of-mass distribution
\cite{Khare:06,Berk-Ladd:07,Winkler-new}.

\subsection{Spectral power for a non-quiescent solvent}
\label{non-quiescent-power-spectrum}

In making use of the above considerations, we now proceed to a prediction of the power spectrum shape
for colloidal particles in an electro-osmotic flow experiment in a closed cell. We follow \cite{TPHV} and use
the convolution integral
 \begin{eqnarray}
\label{eq:convoluted_spectrum}
    C_\textrm{shet}^M({\bf q},f) \;\!=\;\! \int_{v_e + u_S(0)}^{v_e + u_{eo}} \!dv\;\! P(v)
\;\! C^0_\textrm{shet}(\qq,f;v,D_\textrm{eff}) \,,
\end{eqnarray}
as an approximation to the measured power spectrum $C_\textrm{shet}^M({\bf q},f)$. Here,
$C^0_\textrm{shet}(\qq,f)$ is of the form of the single-sphere power spectrum given in Eq. (\ref{eq:powerspectrum-zero}),
with $f = \omega/(2\pi)$.
The integral over the normalized particle velocity distribution function $P(v)$
derived from the Komagata solvent profile extends over a finite velocity
range depicted in Fig. \ref{fig:2}. The velocity-averaged power spectrum
depends also on $u_{eo}$ through its dependence on $P(v)$.
Note that (super)-heterodyne scattering is sensitive to particle velocities taken relative to the laboratory (cell) frame.

We summarize here the assumptions underlying Eq. (\ref{eq:convoluted_spectrum}): a) the particle electrophoretic
velocity is neither coupled to the diffusive motion nor to shear and the stationary parabolic flow profile
is unperturbed by the particle interactions (if present); b) the mean particle density remains unperturbed by the
shearing motion of the solvent and the applied electric field; c) interaction polydispersity and polydispersity in
the electrophoretic velocities are disregarded, and d) density inhomogeneities are not considered
irrespective of structure formation in strongly interacting particle systems.

Furthermore we assume e) on exploiting the dominance of incoherent scattering at low $q$,
that the power spectrum of a homogeneous and quiescent suspension of strongly interacting colloidal spheres
can be described approximately by the single-particle power spectrum $C^0_\textrm{shet}(\qq,f)$ in Eq. (\ref{eq:powerspectrum-zero}).
Here, we make two replacements. First we substitute $D_0$ by $D_\textrm{eff}(\phi)$, where
in the absence of coupling between diffusion and shear we expect that $D_\textrm{eff}(\phi) = D_s(\phi)$,
while at higher shear rates larger values of $D_\textrm{eff}$ may be expected
\cite{Taylor dispersion,Indrani:95,FossBrady:99,Breedveld:03,Leshansky:08}. Second, we replace $v_e^0$ by $v(\phi)$.
Likewise, the 'static' prefactor $\langle I_s^0(\qq) \rangle$ multiplying the relevant super-heterodyne part in Eq. (\ref{eq:powerspectrum-zero}) should be replaced
by $\langle I_s(\qq) \rangle \propto n\;\!f^2(q)\;\!\left[9\;\!s^2 + S(\qq) \right]$ for a weakly size-polydisperse
system.
\begin{figure} \vspace*{0cm}
\epsfig{file=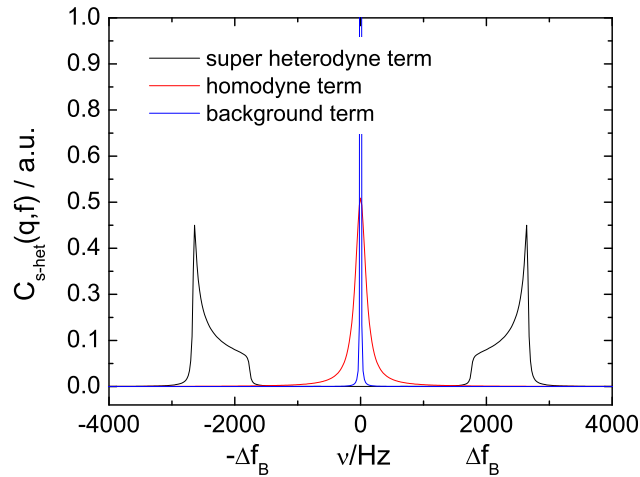,width=9.5cm,angle=0} \caption{\label{fig:5}
Theoretically expected power spectrum according to Eq.
(\ref{eq:convoluted_spectrum}).}
\end{figure}

The power spectrum predicted by Eq. (\ref{eq:convoluted_spectrum}), which is a continuous distribution
 of velocity shifted Lorentzians, is shown in Fig. \ref{fig:5}. It has the typical
super-heterodyne form consisting of two symmetrically displaced heterodyne wings at about
$\pm \Delta f_{B}$, where $\Delta f_B = \Delta\omega_B/(2\pi)$, a sharp zero-frequency background term indicated by the
two narrow vertical lines, and a central homodyne portion symmetrically distributed around zero frequency and unaffected
by the velocity distribution.
The form of the particle velocity distribution depicted in Fig. \ref{fig:2} is
recognized easily in the shape of the  super-heterodyne wings, broadened slightly due to diffusion of
the particles. Experimentally, only the positive-frequency heterodyne wing of the symmetric
power spectrum is resolved, which includes all the information on the velocity distribution and, thanks to the
super-heterodyning, is well separated from the non-interesting homodyne part, and there is no overlap of
the symmetrically displaced super-heterodyne wings of the power spectrum.

For the analysis of the experimental power spectra in terms of the
theoretical spectrum described by Eq.
(\ref{eq:convoluted_spectrum}), the latter is fitted to the
experimental power spectra data (super-heterodyne wing) using four
independent fitting parameters $A$, $v_{e}$, $u_{eo}$ and
$D_\mathrm{eff}$. Since the overall volume flow in a closed cell is
zero so that the mean particle velocity, ${\langle v \rangle}_P$, is
equal to $v_e$, the position of the center of mass of the heterodyne
wing is determined by $v_e$, whereas the osmotic velocity $u_{eo}$
determines its asymmetric broadening due to the solvent flow, and
$D_\mathrm{eff}$ its symmetric broadening due to the assumed
isotropic diffusion. Finally, $A$ is the frequency-integrated part
of the power spectrum arising from the positive-frequency
super-heterodyne wing. By  the sum rule in Eq. (\ref{eq:sum-rule}),
\begin{equation}
\label{eq:integrated-A}
   A \;\! =\;\! I_r \;\! \langle I_s(\qq) \rangle  \,,
\end{equation}
so that $A$ is determined by the prefactor of the super-heterodyne wing.

\section{Experimental results}

Before analyzing the results of our electrophoretic scattering experiments, we
first describe our findings for the field-free static structure factor
of deionized systems. Next we discuss our electrophoretic flow experiments starting first with
suspensions of non-interacting colloidal spheres and progressing then to fluid-like systems of strongly
interacting spheres. Finally, crystalline systems will be shortly addressed.

\subsection{Static light scattering}

\begin{figure}
\centerline{\includegraphics*[angle=0,width=9.5cm]{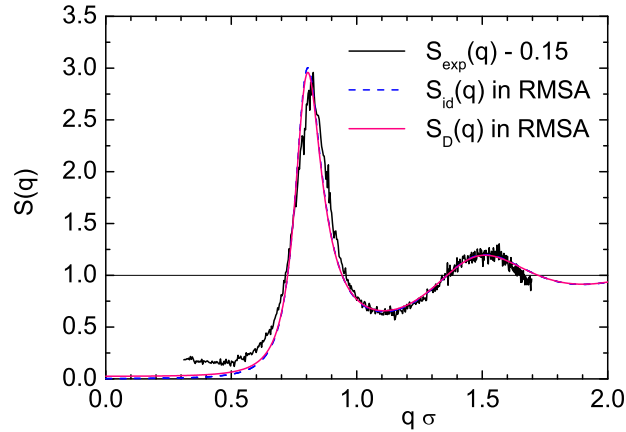}}
\caption{Experimentally determined static structure factor
$S_{exp}(q)$ of a slightly polydisperse, deionized suspension, with
$\sigma = 68$ nm, $\phi = 8.38 \times 10^{-4}$, and $s = 0.05$, in
comparison to the decoupling approximation structure factor
$S_D(q)$, and the structure factor $S_{id}(q)$ of an ideally
monodisperse system calculated in RMSA. The effective particle
charge has been adjusted to match $S_D(q)$ to the peak height of the
experimental $S_{exp}(q)$, by adjusting the effective charge to
$575$ e. Note that $X(q)\approx 9 s^2$ is practically
$q$-independent in the experimental $q$-range. } \label{fig:4}
\end{figure}

A standard static light scattering experiment was employed to obtain the measurable static structure factor, $S_{M}(q)$, from
the division of the (background corrected) scattered mean intensity $\langle I_s(q)\rangle- I_s(q)_{\mathrm{H}_2\mathrm{O}}$ of
an ordered suspension by the (background corrected) scattered intensity from a non-interacting suspension of the same species
(at salt concentration of $c = 4\times10^{-4}\;\! \mathrm{mol/l}$) weighted with the dilution ratio
$n_\textrm{non-interacting}/n_\textrm{ordered}$. The background intensity $I(q)_{\textrm{H}_2\textrm{O}}$ of parasitic stray
light was determined from measurements of $I(q)$ on the pure solvent. The static structure factor shown in Fig. \ref{fig:4} is
for a deionized suspension of PnBAPS68. Experimentally, the pronounced principal peak and the consecutive secondary oscillation
are resolved. From a comparison of the signal obtained for the salty suspension to a Rayleigh-Debye-Gans form factor we found an
angle-independent background intensity of $0.15$ in relative units due to multiple scattering. This offset was subtracted from
the static structure factor measured on the deionized suspension. A fit of the experimental structure factor using the
decoupling approximation of Eq. (\ref{eq:decoupling-Svonq}) for a polydispersity $s = 0.05$, and an ideally one-component $S(q)$
obtained from the rescaled mean-spherical approximation (RMSA) and a DLVO-type screened Coulomb pair potential, describes the
multiple-scattering-corrected data overall quite well. Note that the value of the scattering wave number, $q^\ast$, in our
super-heterodyne flow experiments is in the low-$q$ region of the structure factor, with $q^\ast/q_m \approx 0.4$. From the
theoretical fit, we have extracted the low-$q$ values of $S_D(q=0) \approx 0.03$, and $S(q=0) \approx 0.0034$, respectively,
showing that the incoherent contribution dominates $S_D(q,\tau)$ at very small $q$. A similar theoretical result is found for
the silica particles. In particular for PnBAPS68 the experimental data are above the theoretical expectation, indicating that
there is a larger incoherent contribution. Such a contribution is frequently observed in static light scattering and presumably
results from optical inhomogeneiety of the particles.

\subsection{Electrokinetic flow of non-interacting particles}

At a number density of $n  = 3.95\times 10^{16}\;\! \mathrm{m}^{-3}$ ($\phi \approx 3.3\times 10^{-4}$) and a salt concentration
of $c = 5.15\;\! \mu\mathrm{mol/l}$ corresponding to a reduced screening parameter contribution $\kappa_s a \approx 0.97$ a
suspension of  Si251 spheres is essentially non-interacting. The measurable static structure factor, $S_{M}(q)$, in this system
is practically equal to one. Fig. \ref{fig:6}a shows a series of measured power spectra taken for increasing field strength. The
experimentally accessible frequency range is restricted to values in the vicinity of the Bragg frequency of 2 kHz, whose value
corresponds to a zero Doppler shift for the super-heterodyne part. Here and in the following figures, the frequency $f$ is taken
relative to the Bragg shift, i.e., $f \to f -\Delta f_B$, so that only the Doppler-shifted frequency is displayed. Near the
frequency $f=0$ corresponding to the Bragg shifted value, the data in Fig. \ref{fig:6} are more noisy and some spectra show
additional intensity. This feature is not reproducible and an experimental artefact of still unknown origin. Possibly it stems
from stray light reflections containing some low frequency contaminations superimposing with the reference beam. The overall
spectral shape, however, clearly reflects the diffusion-convoluted velocity distribution of the heterodyne wing depicted in Fig.
\ref{fig:5} and based on Eq. (\ref{eq:convoluted_spectrum}). There is a linear shift of the center of mass of the spectra with
increasing field strength $E$, corresponding to an increasing electrophoretic velocity $|v_{e}| = \mu_e\;\!E$. The width of the
spectra also broadens, as the electro-osmotic velocity, $u_{eo}$, near the cell wall increases linearly with the field strength.
In addition, there also is a symmetric broadening of the spectra due to isotropic diffusion. These features have been
anticipated also in earlier heterodyne dynamic light scattering results \cite{TPHV}, but there they have been masked by homodyne
contributions and overlap with the complementary 'left-handed' heterodyne wing \cite{TPHV}. With our super-heterodyne setup,
these features can be clearly discriminated, allowing thus a fully quantitative evaluation.

Quite interestingly, we find experimentally that the shape of the spectra of non-interacting spheres is given by a single master
curve, in accord with the theoretical expectation when $D_\textrm{eff} \propto E$, $\vv \propto E$ and $P(v) \propto 1/E$. This
can be seen in Fig. \ref{fig:6}b where the spectra superimpose on each other when the scaled intensity $C_\textrm{shet}^M({\bf
q},f)\times E$ is plotted versus the scaled frequency $f/E$.  Except for the spurious contributions of low frequency noise near
the origin, all spectra coincide nicely within the experimental errors. This shows that the simple superposition of
velocity-weighted single-particle Lorentzians used in Eq. ({\ref{eq:convoluted_spectrum}) works quite well for the considered
systems of non-correlated particles, where $S_M(q) \approx 1$. In other words the measured spectra can be directly interpreted
as velocity distributions broadened by diffusion-like mechanisms characterized by $D_\textrm{eff}$. This allows us to include a
velocity scale, at the top of Fig. \ref{fig:6}a by multiplying $f$ with $2\pi/q =0.796$.

\begin{figure} \vspace*{0cm}
\epsfig{file=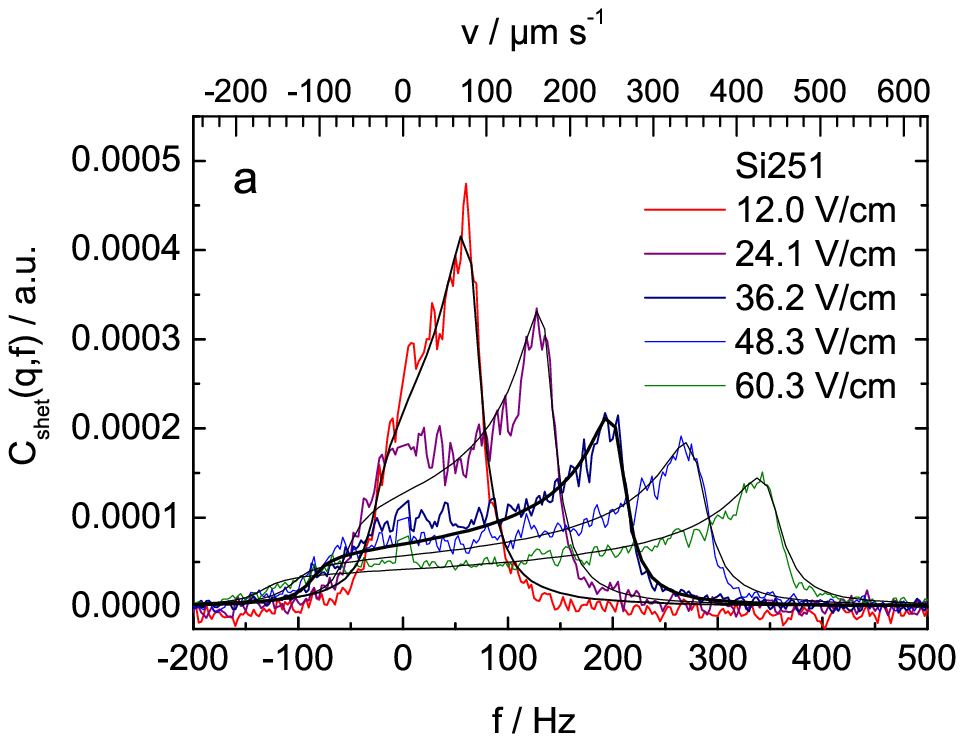,width=8.0cm,angle=0}
\epsfig{file=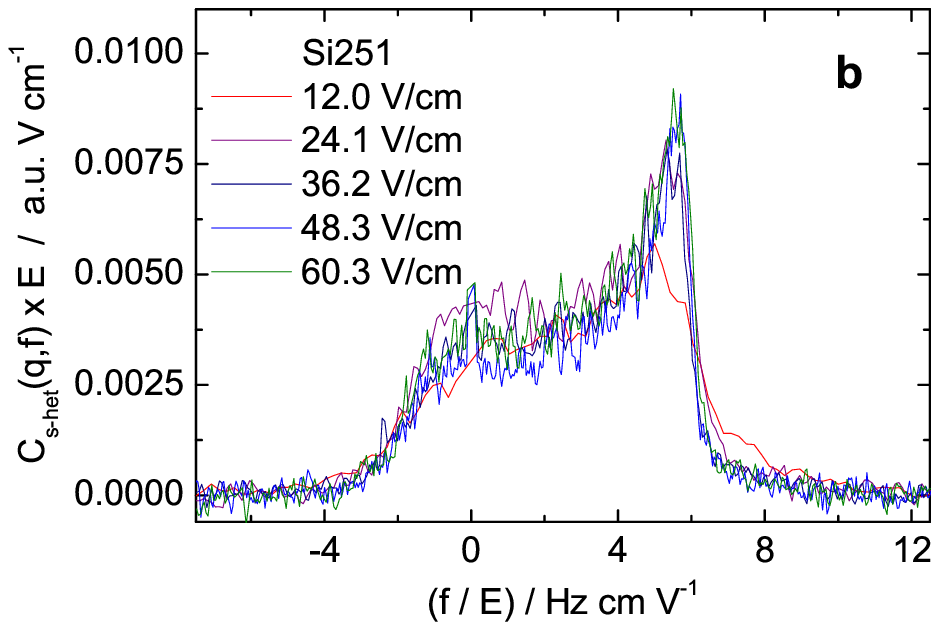,width=8.0cm,angle=0}
 \caption{\label{fig:6}
Power spectra obtained for non-interacting Si251 spheres at a salt concentration of $c = 5.15\;\!\mu \mathrm{mol/l}$ and a number density of $n  = 3.95 \times 10^{16}\;\! \mathrm{m}^{-3}$. The
frequency is taken relative to the Bragg-shift $\Delta f_B$. The smooth solid lines are fits of Eq. \ref{eq:convoluted_spectrum} to the data. The power spectra are well described in all cases,
thus the superposition procedure works well. The upper scale shows the corresponding velocities obtained from $v = 2\;\!\pi\;\!f/q$. b: scaled spectral power versus reduced frequency. All
spectra nicely coincide, showing that in all cases the flow is parabolic.}
\end{figure}

\begin{figure} \vspace*{0cm}
\epsfig{file=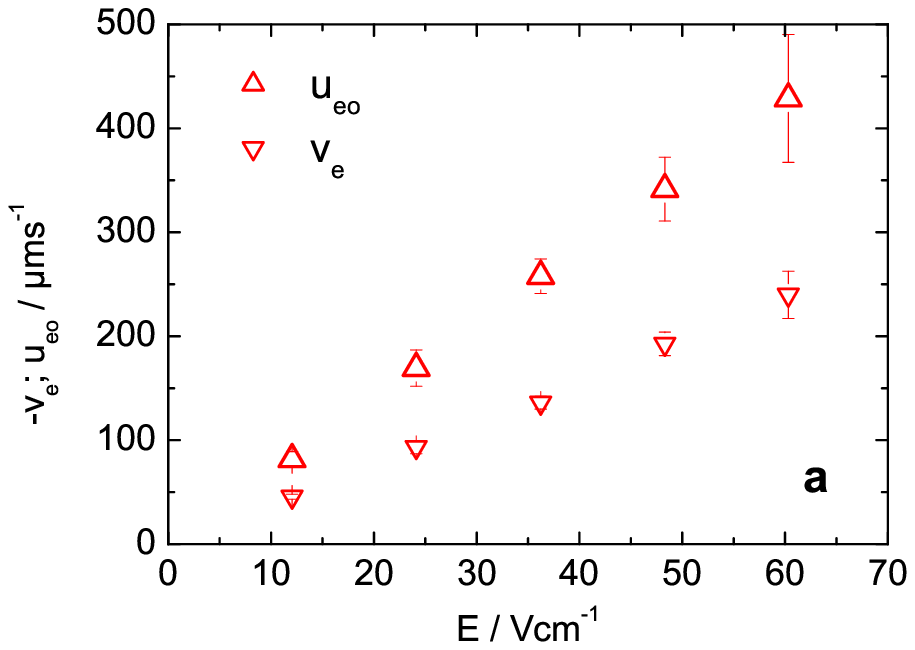,width=7.5cm,angle=0} \epsfig{file=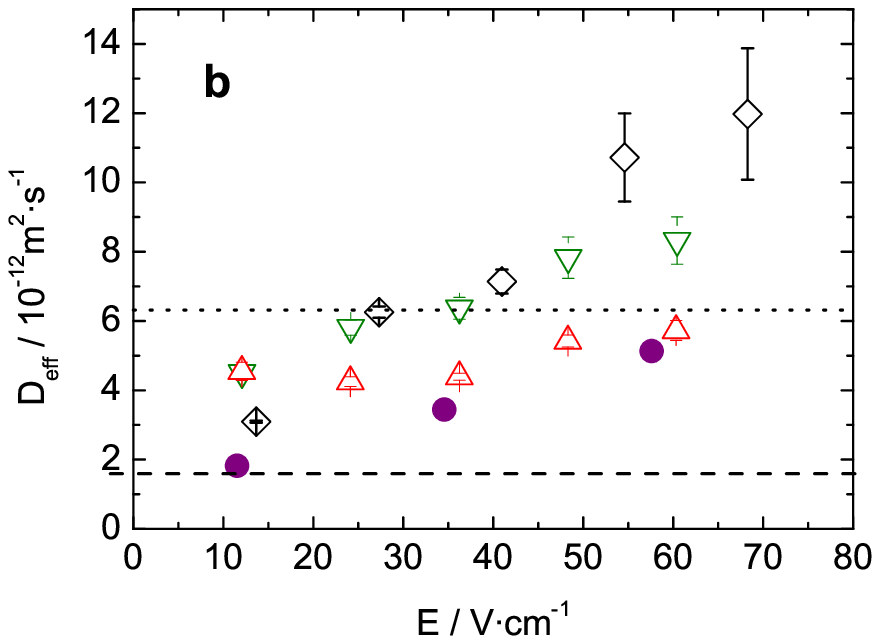,width=7.5cm,angle=0}
\epsfig{file=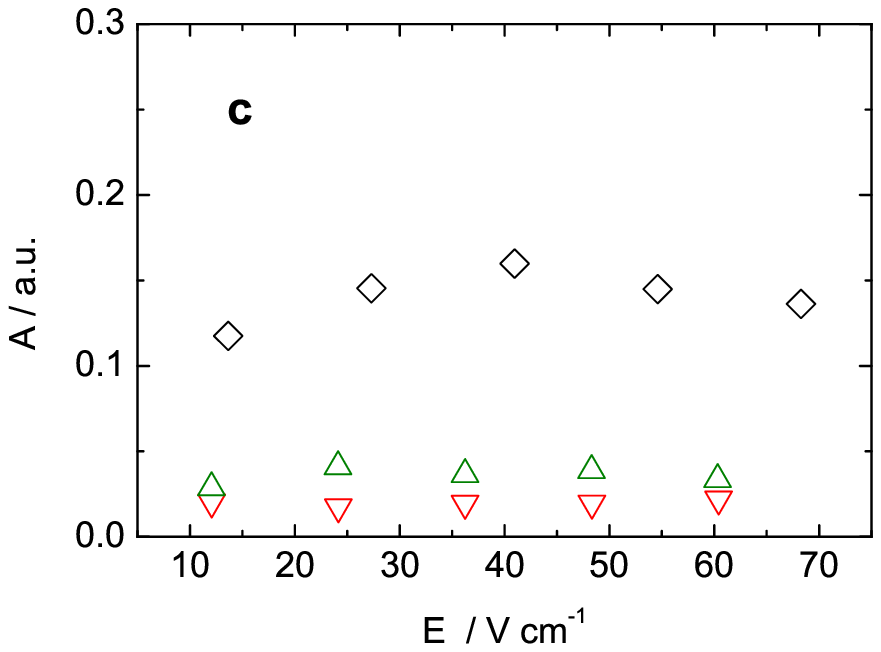,width=7.5cm,angle=0}\caption{\label{fig:7} Fit parameter results in dependence on the applied field
strength. a: electrokinetic velocities, $|v_e|$ (down triangles) and $u_{eo}$ (up triangles) for non-interacting Si251 at $n =
3.95\times 10^{16}\;\! \mathrm{m}^{-3}$ and $c = 1.32 \;\! \mu\mathrm{mol/l}$; b: comparison of effective diffusion constants,
$D_\textrm{eff}$, for the following suspensions: Si251 at $n = 3.95\times 10^{16}\;\! \mathrm{m}^{-3}$ and $c = 5.15 \;\!
\mu\mathrm{mol/l}$ (up triangles), Si251 at $n = 3.95\times 10^{16}\;\! \mathrm{m}^{-3}$ and $c = 1.32\; \mu\mathrm{mol/l}$
(down triangles), Si251 at $n = 3.95\times 10^{16}\;\! \mathrm{m}^{-3}$ and $c = 1.07\;\! \mu\mathrm{mol/l}$ (diamonds) and
PnBAPS68 at $n = 0.47\;\! \mu\mathrm{m}^{-3}$ and $c = 0.2 \;\! \mu \mathrm{mol/l}$ (closed circles). The two horizontal lines
indicate the Stokes-Einstein values $D_0 = 6.30 \times 10^{-12}\;\!\mathrm{m}^2/\mathrm{s}$ and $D_0 = 1.71 \times
10^{-12}\;\!\mathrm{m}^2/\mathrm{s}$ for PnBAPS68 (dotted) and Si251 (dashed), respectively. c: comparison of integrated
spectral intensities $A$ for Si251. Symbols as before. The principal peak of $S(q)$ in the three suspensions takes values of
about $1$, $1.3$ and $1.7$, respectively, indicating increasing particle correlations. }
\end{figure}

We have fitted all measured spectra by Eq. (\ref{eq:convoluted_spectrum}) using  the four independently adjustable parameters:
$v_{e}$, $u_{eo}$, $D_\mathrm{eff}$ and integrated scattering signal strength $A$. The fits describe the experimental power
spectra of non-interacting particles quite well. This implies, as expected, that the flow profile in these systems is of
parabolic shape. The results for the optimized parameters of these fits are included in Figs. \ref{fig:7}a-c and compared with
the results for suspensions of strongly interacting particles that will be discussed in the following subsection.  Both the
electrophoretic and electro-osmotic velocities deduced from the fit increase linearly with increasing field strength (see Fig.
\ref{fig:7}a). This is an important first check, showing that we are in the linear response limit of electrokinetics and that
the applied field strength is much smaller than the field within a particle double layer. Our fits further yield $u_{eo} >
|v_{e}|$, corresponding to a strong electro-osmotic shear flow close to the wall.

Furthermore, a roughly field-independent effective diffusion coefficient is obtained for the non-interacting case with
$D_\mathrm{eff} = 4\times10^{-12}\;\! \mathrm{m}^{2}/\mathrm{s}$. The coefficient has a value somewhat larger than the
Stokes-Einstein value $D_0 \approx 1.7 \times 10^{-12}\;\! \mathrm{m}^{2}/\mathrm{s}$ of the Si251 spheres, and shows the
indication of a slight increase with the applied field. A more strongly pronounced and linear increase of $D_\textrm{eff}$ with
the applied field has been observed in an earlier study on low salt, but non-interacting polystyrene latex suspensions
\cite{TPHV}, where $D_\textrm{eff}$ was found to extrapolate approximately to $D_0$ in the zero-field limit. The enlarged value
for $D_\textrm{eff}$ at small $E$ may have several origins. Most probable, transient time broadening, and/or an increased sample
temperature due to Joule heating expected at this elevated salt concentration cause the enlarged $D_{\textrm{eff}}$. The
possible additional increase with increasing field will be discussed in detail below.

The frequency-integrated power spectrum of the non-interacting system is found to be essentially field-independent (up triangles
in Fig. \ref{fig:7}c). According to Eq. (\ref{eq:integrated-A}), the frequency-integrated intensity of the super-heterodyne wing
for non-interacting monodisperse particles is given by $A \propto n\;\! P(q)$, where $P(q)$ is the single-sphere form factor. It
is independent of the (normalized) particle velocity distribution and $D_\textrm{eff}$. An enlarged $E$ stretches the
super-heterodyne wing due to the increased $u_{eo}$ (solvent flow profile becomes more parabolically stretched out), and
frequency-shifts its center of mass due to the increased value of $v_e$, but the frequency-integrated super-heterodyne part,
i.e., the 'area' $A$ under the wing, should remain constant unless $P(q)$ or $n$ would change significantly in the presence of a
field.

\subsection{Electrokinetic flow of fluid-like suspensions of interacting particles}

We now proceed from non-interacting
suspensions to systems of increasingly pronounced short-range order
and decreasing isothermal osmotic compressibility. Upon reducing the salt
concentration at constant $n  = 3.95\times 10^{16}\;\!
\mathrm{m}^{-3}$ for Si251, or increasing the particle
concentration of deionized ($c = 0.2 \;\! \mu \mathrm{mol/l}$)
PnBAPS68 spheres from $n  = 0.47\;\! \mu\mathrm{m}^{-3}$ to $n  =
4.7\;\! \mu\mathrm{m}^{-3}$, the peak heights of the field-free
static structure factor have values of $1.3$ and $1.7$ for
Si251, and $2.0$ and $2.6$ for PnBAPS68. Thus all
these samples are in the isotropic fluid state with their
short range order becoming more and more pronounced in that order.
We first discuss the spectral shapes and then turn to the effective
diffusivities and the frequency-integrated power spectrum.

\begin{figure} \vspace*{0cm}
\epsfig{file=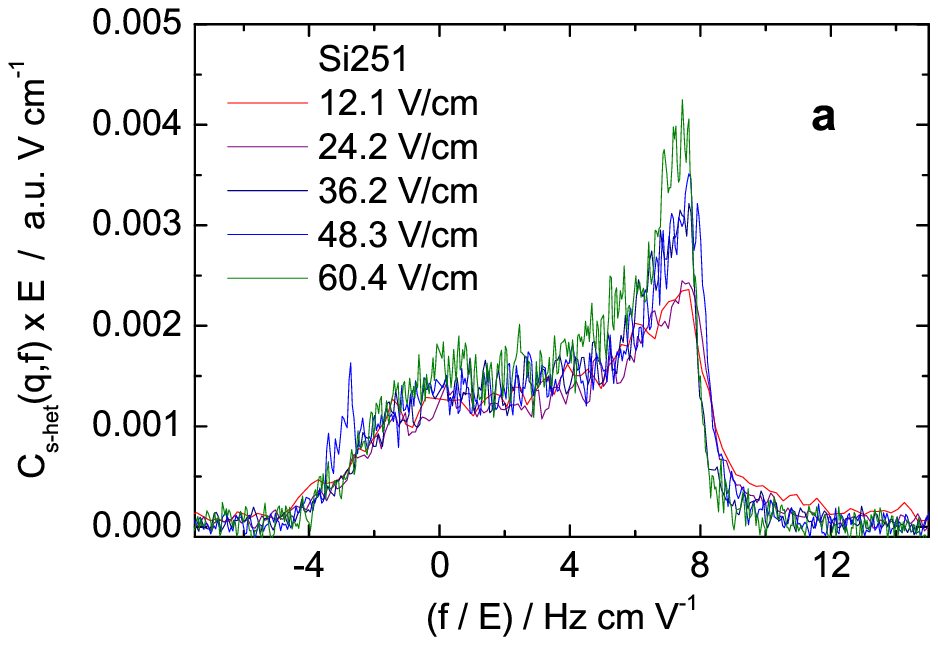,width=7.5cm,angle=0}
\epsfig{file=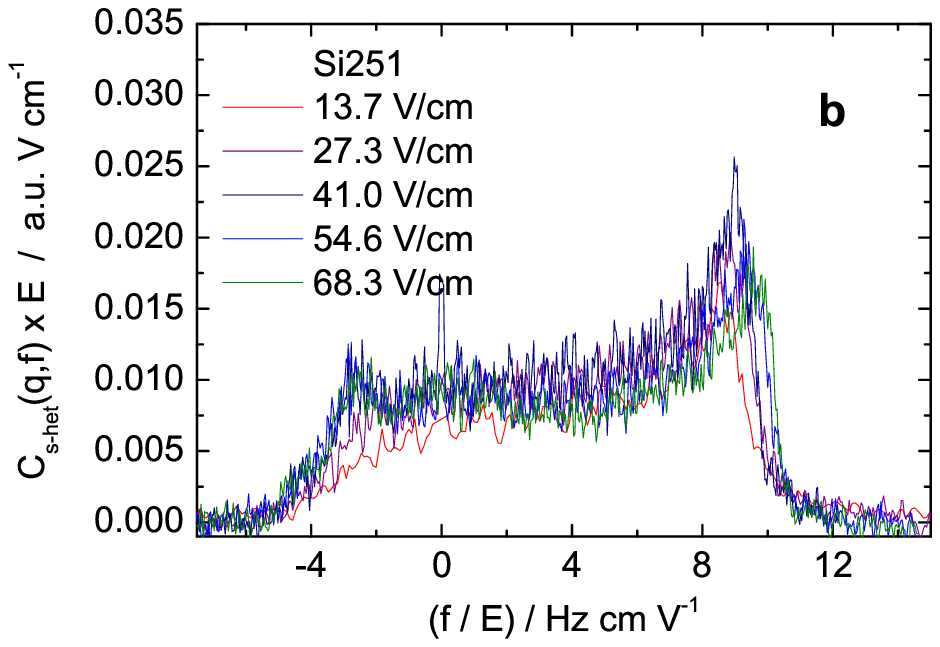,width=7.5cm,angle=0}
\epsfig{file=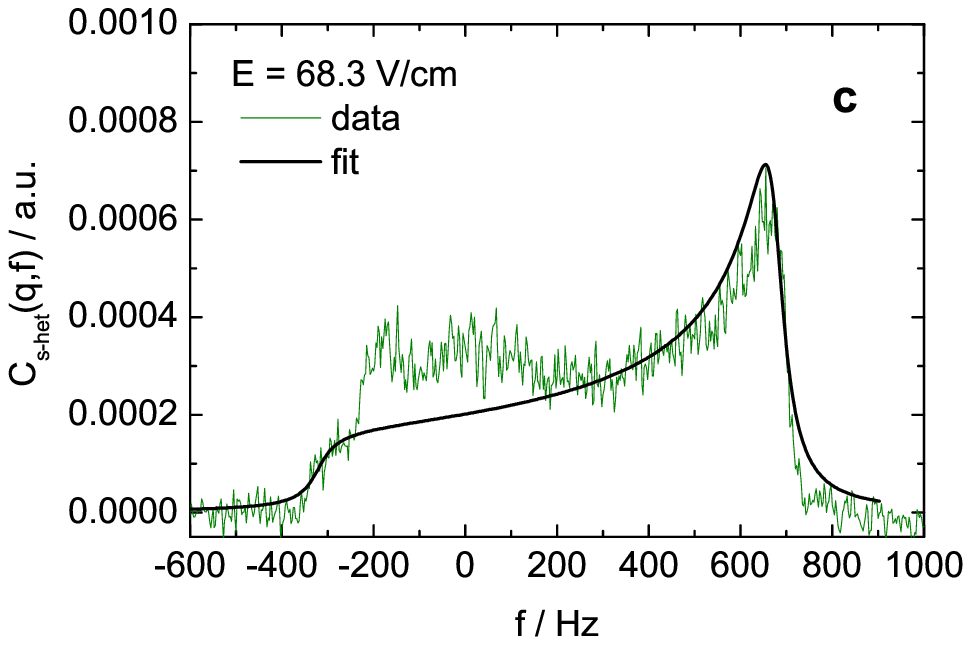,width=7.5cm,angle=0} \caption{\label{fig:8}
Suspensions of weakly and moderately strongly correlated particles
of Si251 at $n  = 3.95 \times 10^{16}\;\! \mathrm{m}^{-3}$. a: field-strength
scaled spectra at $c = 1.32 \;\!\mu \mathrm{mol/l}$, for
field strengths as indicated; b: as in a but for $c = 1.07 \;\!
\mu\mathrm{mol/l}$; c: comparison of the $E = 68.3$ V/cm spectrum in
b: with the fit curve derived from Eq. (\ref{eq:convoluted_spectrum}). Note the
extra spectral power in the experimental data at moderately small
frequencies. }
\end{figure}

Let us first make a general remark on spectra recorded at moderately strong interactions. According to the sum rule in Eq.
(\ref{eq:self-part-scattering}), $g_E(\qq,\tau) \;\! \approx \;\! {|c|}^2 \;\! N\;\! f^2(q) \left [ 9 \;\! s^2 \;\! G(\qq,\tau) +
S(\qq,\tau) \right] \,,$ the IACF is a weighted superposition of coherent and an incoherent part. An expression for the coherent
part in an external field is not yet known. Correspondingly, the integrated spectral power in Eq. (\ref{eq:integrated-A}) is given by $A \propto
n\;\!P(q)\;\!\left[9\;\!s^2 + S(\qq) \right]$ for slightly size-polydisperse systems and, more generally,
by $A \propto n\;\!
\overline{f^2}(q)\;\!S_M(\qq)$. With increasing interactions the low-$q$ form of $S(\qq)$ decreases but
the the incoherent scattering contribution may not yet dominate the coherent one. In such a situation, the  major assumption underlying Eq. (\ref{eq:self-part-only}), namely the dominance of the incoherent
scattering part, does not necessarily apply. Rather, we expect also the coherent
(steady-state) dynamic structure factor, $S(\qq,\tau)$, to
contribute to the spectrum. And since to date no exact theoretical expression is known for this quantity and its Fourier transform,
one should be cautious in interpreting spectra of systems of moderately interacting particles as velocity distributions. On the
other hand, any (field-dependent) deviation from the expected spectral shape and amplitude may indicate interesting underlying
physics.

In Fig. \ref{fig:8}a, we show the field-scaled power spectra for the weakly interacting silica spheres. Again all spectra
superimpose neatly. Only minor changes of the spectral shape are observed, mainly appearing as a slight increase in the peak
region as the field is increased. In Fig. \ref{fig:8}b, five spectra are superimposed for the moderately strongly interacting
silica suspension. The field-dependent changes of the spectral shape in Fig. 8b are more pronounced than before. In particular one
observes a relative increase of spectral power for the high-shear, low velocity part of the spectra. While this feature is
essentially absent at the lowest fields and high-quality fits are obtained, the fits at large fields are considerably worse.
This is highlighted in Fig. \ref{fig:8}c, where we compare the (unscaled) experimental super-heterodyne signal to a fit of Eq.
(\ref{eq:convoluted_spectrum}). The difference between expected and measured spectral power distribution in the interval
$-200\;\! \mathrm{Hz} < f < 200$ Hz is clearly visible. Neglecting this region, a reasonably accurate description of the data is
still achieved, compromising between peak intensity and diffusive-like broadening. A fit by Eq. (\ref{eq:convoluted_spectrum}) used for
is based on the assumption of homogeneously distributed, identical scatterers and a parabolic flow profile. Thus either local changes of
scattered intensity or deviations from parabolic flow may cause the deviations observed. The restricted frequency range for the
relative enhancement of the experimental spectral power points to a local effect. One possible cause could be shear- or wall-induced changes in
$S(\qq,\tau)$. In principle, it should be even possible to discriminate between local effects on the scattering power and global
changes of the flow profile. To this end we checked for the linearity of the increase in the velocity corresponding to the center of
mass of the spectra with increasing field strength. An increased scattering power at low frequencies would indicate a decreased apparent
velocity. In contrast to this, a changed flow profile would no be such an indication, since for closed cell boundary conditions the integrated solvent
motion must average to zero. In the present case, the observed deviations from a linear increase of the velocity with increasing field was still below the experimental uncertainty in $v_e$.

\begin{figure} \vspace*{0cm}
\epsfig{file=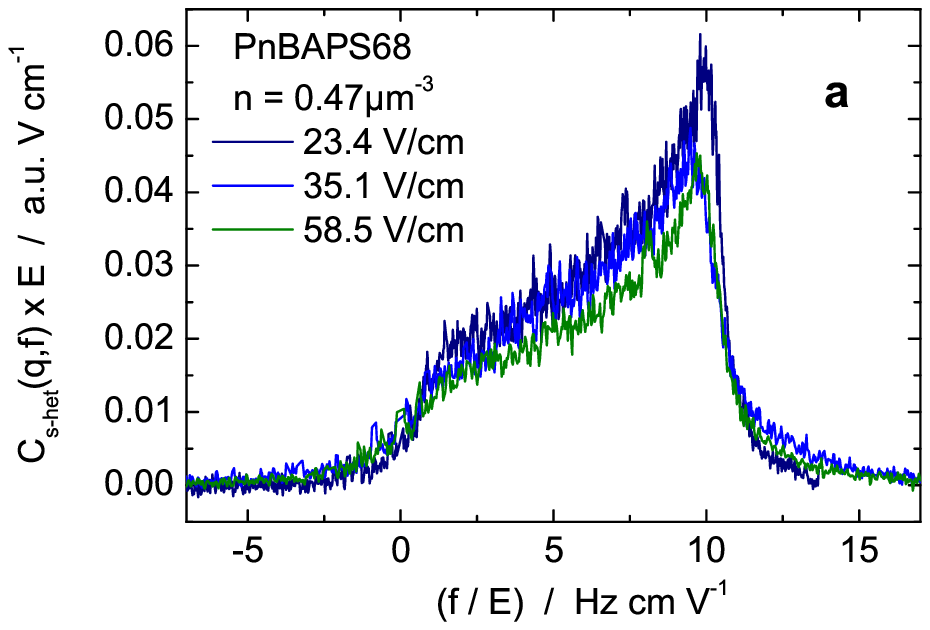,width=7.5cm,angle=0} \epsfig{file=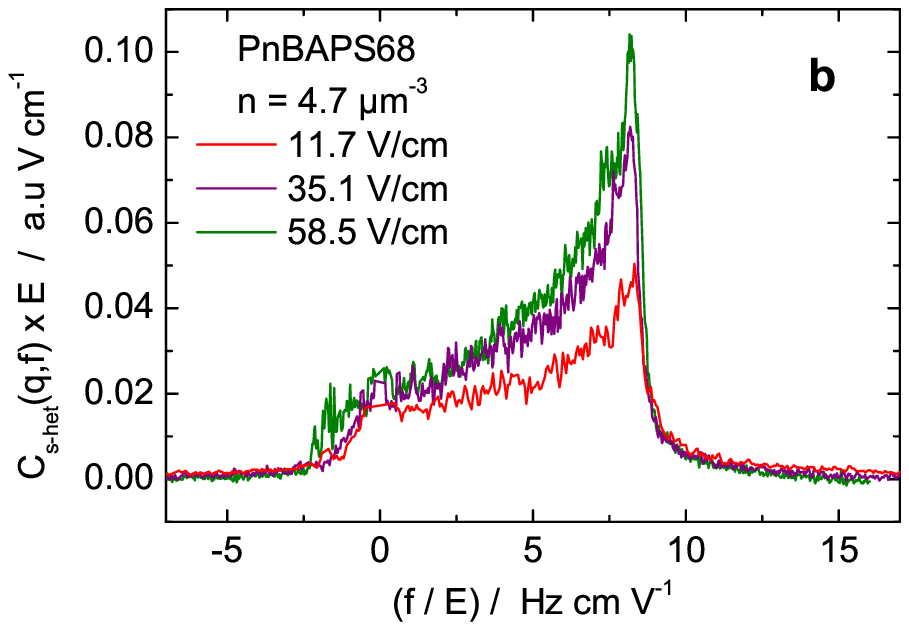,width=7.5cm,angle=0}
\epsfig{file=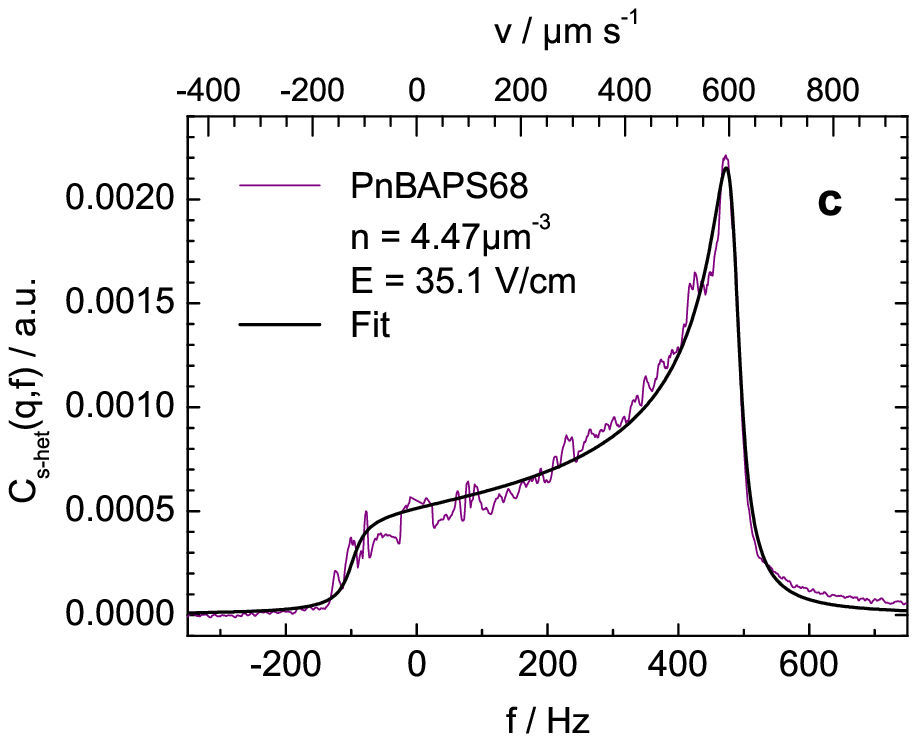,width=7.5cm,angle=0} \epsfig{file=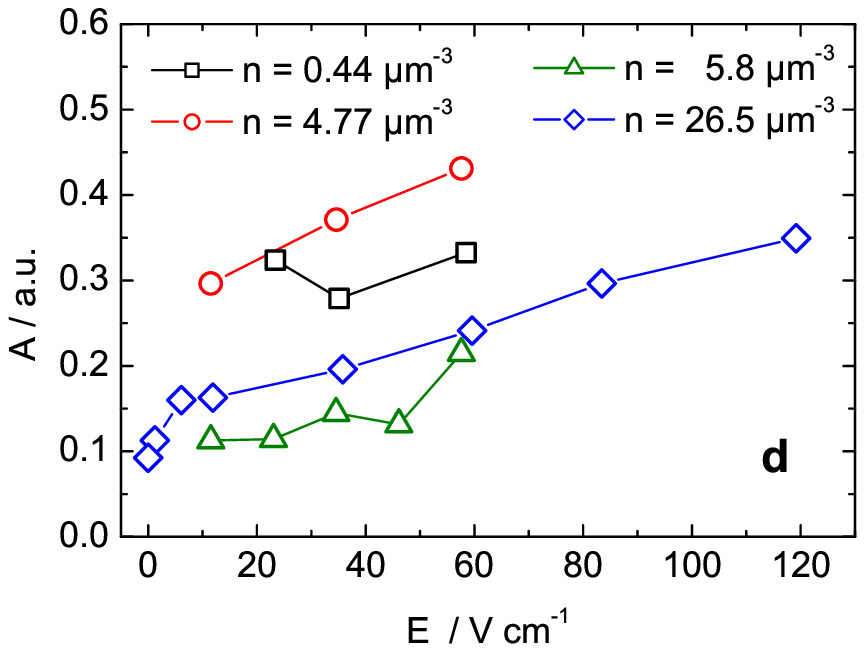,width=7.5cm,angle=0} \caption{ \label{fig:9} Fluid-like
suspensions of strongly interacting PnBAPS68 particles. a: scaled spectra for PnBAPS68 at $n  = 0.47\;\! \mu\mathrm{m}^{-3}$ and
$c = 0.2 \;\! \mu \mathrm{mol/l}$, for field strengths as indicated; b: scaled spectra at $n  = 4.7\;\! \mu\mathrm{m}^{-3}$ and
$c = 0.2 \mu \mathrm{mol/l}$ for field strengths as indicated. c: background corrected spectrum for $n  = 4.7\;\!
\mu\mathrm{m}^{-3}$ and $E = 35.1V/cm$, with fit curve according to Eq. (\ref{eq:convoluted_spectrum}) d: field-strength dependence of the
integrated spectral power for different particle concentrations as indicated. }
\end{figure}

\begin{figure} \vspace*{0cm}
\epsfig{file=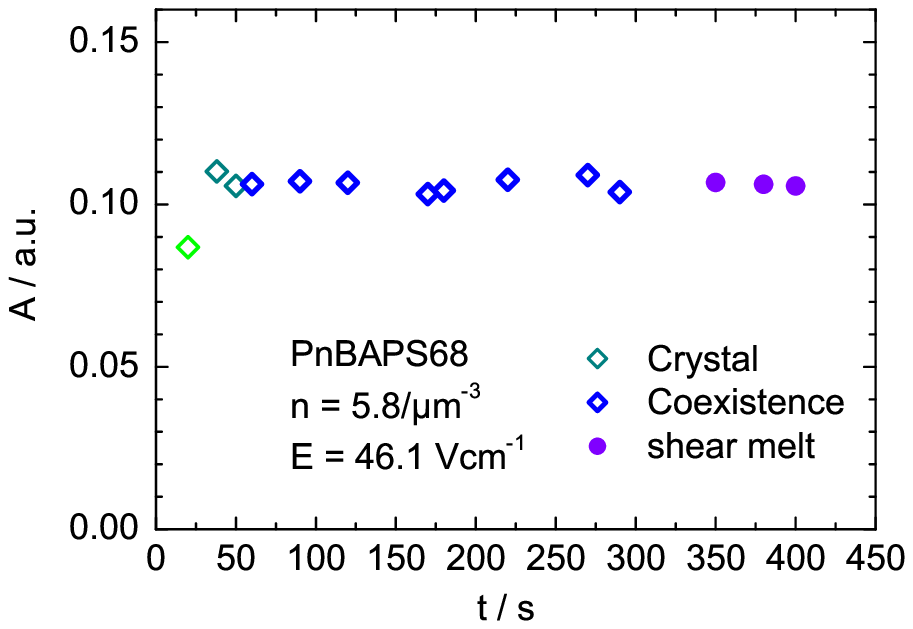,width=9.5cm,angle=0} \caption{\label{fig:10} Integrated spectral power for a shear melting suspension of
PnBAPS68 at $n = 5.8\;\! \mu\mathrm{m}^{-3}$. A field of $E = 46.1$ V/cm was applied at $t=25$ s. After the initial jump from
its zero-field value, $A$ stays constant irrespective of the system structure}.
\end{figure}

In Figs. \ref{fig:9}a-b, we show the scaled spectra for low-salinity suspensions of strongly interacting PnBAPS68 spheres. The
extension of the spectra in terms of scaled frequency is more narrow than in the spectra
considered before, and they start at about zero frequency. The
electro-osmotic velocity in this case is much smaller, and on the order of the particle electrophoretic velocity. In turn, this
implies a less curved parabolic solvent profile stretching to smaller velocities in the cell center. Interestingly, the low-$f$
overshoot of the experimental power spectrum is absent in these samples. Second, in particular for the more concentrated sample,
the overall spectral power increases significantly with the applied field. Still, all these spectra can be fitted rather well
using Eq. (\ref{eq:convoluted_spectrum}), and Fig. 9c gives an example of the excellent fit quality. The simple superposition
procedure in Eq. (\ref{eq:convoluted_spectrum}) again works well and the parapolic flow profile is confirmed. Here the reduction
of the low-$q$ form of $S({\bf q})$ is sufficient to obtain a super-heterodyne spectrum dominated by incoherently
scattered light, in accord with the requirement underlying Eq. (\ref{eq:self-part-only}). Thus, like for the non-interacting
case, we again may interpret these spectra as velocity distributions. This main result of the experimental part of our paper on
flow measurements is corroborated by the observation that the experimentally determined electrophoretic velocities are in
excellent agreement with both state-of-the-art analytical calculations \cite{Carrique}, and combined Monte Carlo and
Lattice-Boltzmann computer simulations on the Primitive Model level \cite{ELS-PRL, Lobaskin:04}.

Given this, a further comparison of the other two fit parameters reveals some interesting deviations which seem to contain
information on interesting underlying physics, not yet fully understood. The effective diffusion coefficients, $D_\textrm{eff}$,
of the silica systems at different salinities and as a function of increased field were compared in Fig. \ref{fig:7}b to those
of PnBAPS at $n=0.47µm^{-3}$. For the three Si251 samples, the zero-field extrapolated values are above the Stokes-Einstein
result (dashed line) with a mild decrease with increasing particle interactions. It is well known that the (equilibrium) long-time self-diffusion
coefficient decreases below the Stokes-Einstein value with increasing interactions, whereas the opposite trend appears for the
collective diffusion coefficient $D_c = D_0\;\!H(0)/S(0)$ \cite{Pusey}. Our finding therefore nicely illustrates the sum
rule in Eq. (\ref{eq:self-part-scattering}), as it shows an originally dominating, but decreasing influence of the coherently
scattered light with decreasing isothermal compressibility, which in turn lowers the contribution of collective diffusion. For
the deionized suspension of PnBAPS68 spheres, linear extrapolation gives a value $D_\textrm{eff}(E=0) \approx 1 \times
10^{-12}\;\! \mathrm{m}^2/\mathrm{s}$ well below $D_0$ (dotted line), and in accord with the expectation for the ´value of the long-time
self-diffusion coefficient. Thus, estimates of the latter quantity may be obtained from extrapolation of $D_\textrm{eff}(E=0)$ to
zero field, but only for strongly interacting fluid systems, where incoherent scattering dominates and additional broadening
mechanisms are absent.

Fig. 7b further reveals an increase of the effective diffusion coefficient, $D_\textrm{eff}$, with increasing field strength. The effect is
only slightly visible at low $n$ and large salt concentrations, but becomes more pronounced for decreased salinity in Si251, and
increased particle number density in PnBAPS68. The field dependence thus increases with increasing  interaction strength.
Conversely, the approximate theory embodied in Eq. (\ref{eq:convoluted_spectrum}) predicts a field strength independent
diffusivity, which in fact was observed in the non-interacting case. In the deionized case, Joule heating cannot be responsible
for this dependence, as it vanishes with decreasing salt concentration. Transient time broadening, which couples the particle
velocity to a non-diffusive peak-broadening, may in principle play a role, but the experimentally observed broadening is too
large for this \cite{TPHV}. In addition, the effect is also pronouncedly present for PnBAPS68 at strong fluid-like order, where
the electro-osmotic profile, and thus the absolute velocities are much smaller. We therefore exclude both mechanisms from
causing the observed field dependence of the effective diffusion coefficient. A number of possible causes will be discussed at
the end of this sub-section.

The field dependence of the integrated spectral power is shown in Fig. 7c for the three silica systems and in Fig. 9d for four
differently concentrated suspensions of PnBAPS68. The integrated power for the silica systems does show roughly field
independent values, which differ for each suspension. Also for PnBAPS68 each system shows a different overall intensity but here
for the more concentrated systems a trend of $A$ to increase with the applied field is obvious. The first observation is traced
back to a detail of the experimental procedure. Each system has a different scattering power at low $q$. This is either due to
the different isothermal compressibility (for Si251 at different salinity) or different particle concentration (PnBAPS68). With
changing scattering power of the suspension, also the reference beam intensity has been adjusted to obtain an optimized signal-to-noise ratio. This manual adjustment, however, is not precisely defined, leading to differences in $A$ for different suspensions.
Hence the absolute values for $A$ should not be compared. For a given suspension, however, the optical adjustment is kept and
hence the field dependence can be studied without ambiguity. Therefore the observation of an increased spectral power with increased field
strength is an interesting finding, which is possibly connected to the increase of the effective diffusion coefficient with the applied
field. Recalling that in general, $A \propto n\;\! \overline{f^2}(q)\;\!S_M(\qq)$, a field induced density fluctuation on a
suitable length scale will lead to an enlarged value of $A$ with increased field. To pursue this idea,  further, we performed a
measurement of the frequency-integrated power spectrum as a function of time for a shear-melting suspension of PnBAPS68 at $n =
5.8\;\! \mu\mathrm{m}^{-3}$ with the spectra shown in Fig. 10. The integrated spectral power jumped discontinuously from its zero-field
value to a 30\% larger value upon application of the field. The fully crystalline state was stable for about half a minute, then the
crystal slowly melted inwards and the final stationary state of a shear-molten structure in the mid-plane of the cell was reached
after some five minutes. During the whole process, $A$ stayed practically constant. This suggests that the increase of spectral
power observed in the field-dependent measurements is not due to a change in the structure. Rather, the initial jump seems to
indicate a correlation with the presence of the electric field.

Several previous studies were concerned with the coupling of diffusivity and/or density fluctuations to external fields. Qiu et
al. \cite{FRS diffusion unter shear} studied the long-time self-diffusion of non- or weakly interacting charged spheres under
low frequency, oscillatory Couette flow. With increasing shear, they observed anisotropic self-diffusion, with the mean-squared
displacement in the direction of the flow field growing faster than linear in time. This was attributed to Taylor dispersion
driven by the position-dependent flow \cite{Taylor dispersion}. In addition, a non-convective enhancement of diffusion was found
also perpendicular to the flow direction, characterized by a diffusion constant that increases with the shear rate. This
shear-induced enhancement of self-diffusion in strongly sheared, dilute suspensions of charged particles was attributed by
Breedveld and Levine to the electrostatic breaking of the time reversal symmetry of particle trajectories in low Reynolds number
flow \cite{Breedveld:03}. Stokesian dynamics simulations for non-dilute, sheared suspensions of hard spheres also show a
pronounced shear-induced enhancement of long-time self-diffusion perpendicular to the flow direction \cite{FossBrady:99}.
Conversely, recent analytic calculations and Brownian dynamics simulations (without hydrodynamic interactions) of collective diffusion of hard spheres
\cite{Leshansky:08} found the shear-induced change in the collective diffusivity tensor away from isotropy to be quite small at
small shear rates (Peclet numbers). Not much is as yet known about suspensions of strongly interacting particles under shear. But, due to the
overall parabolic flow profile for fluid-ordered suspensions in the present experiments, we expect an inhomogeneous spatial
distribution of observable effects.

In a recent study on the electrokinetics of charged spheres in a quiescent solvent, Araki and Tanaka \cite{Tanaka:08} performed
numerical calculations based on a so-called fluid particle dynamics (FPD) method, where the colloidal particles are described as
highly viscous liquid spheres and the solvent and microions are treated as continuous fields akin to standard electrokinetic
theory. The FPD method accounts for many-particle hydrodynamic interactions. Within the linear response regime of the
field-dependence of the electrokinetic velocities, the authors found hydrodynamically induced velocity fluctuations which
increased with the field strength. Clearly, such fluctuations will directly mimic an enhanced diffusivity. For low field
strengths, the velocity fluctuations decreased monotonically with increasing salinity but at larger field strength the scaled
velocity fluctuations pass through a minimum at some intermediate salinity value. The authors further pointed out that these
fluctuations significantly alter the pair correlation function, i.e. they translated into density fluctuations. It will be
interesting to explore how such fluctuations, which should likewise grow with the electric field, behave on large spatial scales
and translate into alterations of the low-$q$ $S(\qq)$.

Finally, an enlarged apparent diffusivity and corresponding global density fluctuations of non-hydrodynamic origin, can be
expected when fluctuations in relative particle velocities are caused by a non-negligible mobility polydispersity. Here
particles move at different constant speeds which introduces a broadening of the velocity distribution. Similar to hydrodynamic
broadening, this effect is independent of the absolute particle velocity but proportional to the relative velocity between
neighboring particles $i$ and $j$ which will increase with the applied field: $\Delta v=(\mu_i - \mu_j )E$. In the former case
the effective diffusivity is enhanced, in the latter the velocity distribution is convoluted with the mobility distribution
(e.g., of Gaussian shape for a narrow distribution). As both types of independent spectral broadening mechanisms cannot be
discriminated properly within the spectra, polydispersity of mobilities may mimic an increased diffusion. On the other hand,
also the suspension microstructure is likely to be affected by the systematic differential motion of neighboring particles. If
density fluctuations occur at the appropriate scale, the relative amplitude of the coherent scattering contribution would be
enhanced. This spatially global effect should become stronger with increasing field strength, but could also decrease with
increasing particle interactions, as the isothermal osmotic compressibility decreases and the structural ordering becomes more
pronounced. A discrimination between all these possibilities may be attempted by applying our method to different flow
situations, with particular focus on those, where the external field couples differently to particles, micro-ions and solvent.

\subsection{Electrokinetic flow of crystalline systems}

\begin{figure} \vspace*{0cm}
\epsfig{file=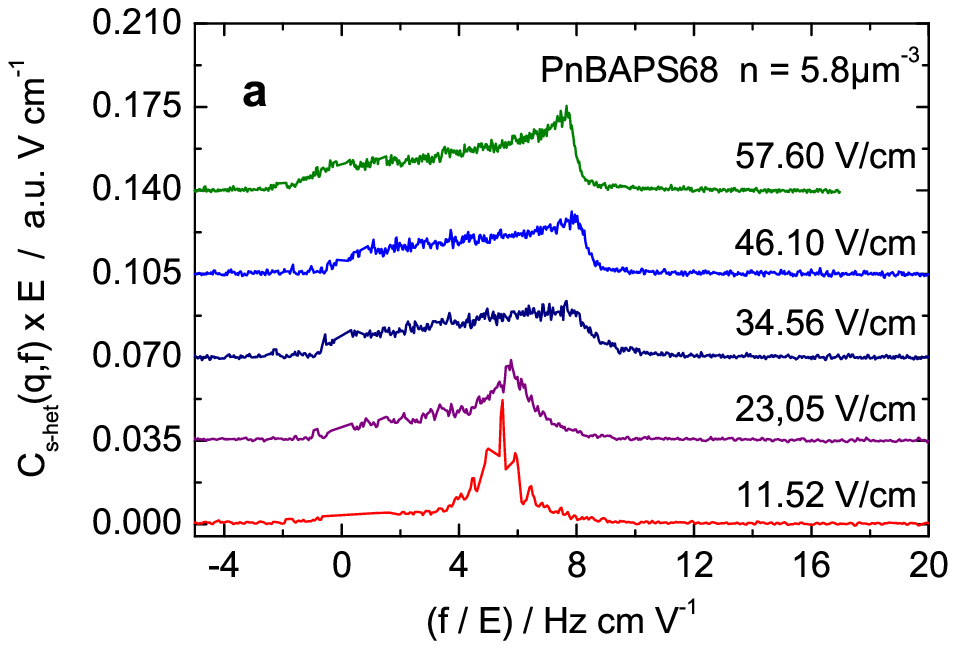,width=7.5cm,angle=0} \epsfig{file=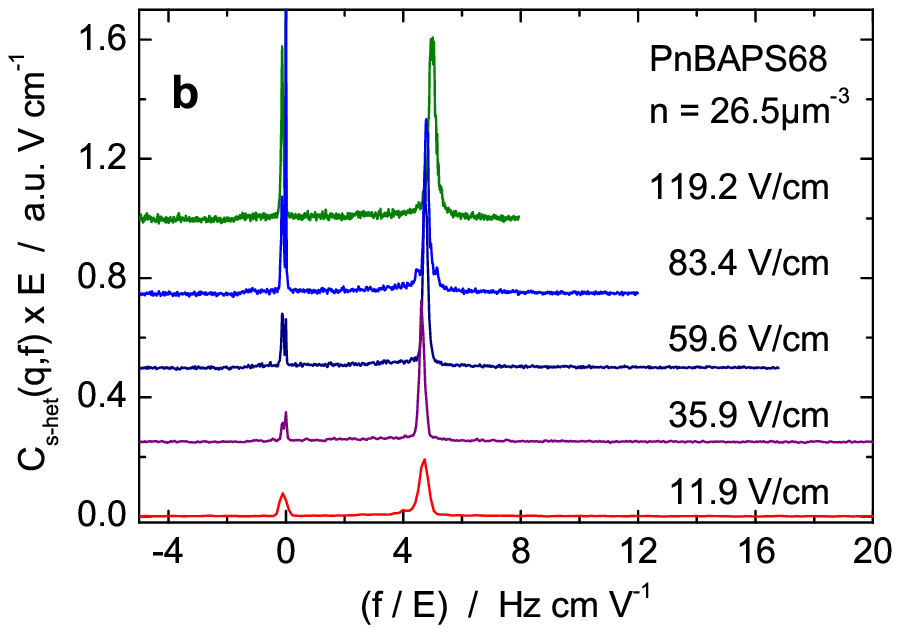,width=7.5cm,angle=0} \caption{\label{fig:11} Scaled power
spectra for two equilibrium crystalline suspensions of PnBAPS68. Spectra are shifted upward for clarity. a: $n = 5.8\;\!
\mu\mathrm{m}^{-3}$. With increasing field strength this system consecutively shows double plug flow, layer crystal plastic flow and complete
shear melting; b: $n = 5.8\;\! \mu\mathrm{m}^{-3}$. With increasing field strength this system shows double plug flow but no
shear melting. All flow profiles, even those in the shear molten state, deviate strongly from the parabolic flow profile
entering into Eq. (\ref{eq:convoluted_spectrum}) .}
\end{figure}

In this last subsection, we shortly address crystalline systems as formed in deionized suspensions of PnBAPS68. In this case, the
finite shear moduli and the different viscosities of the different phases coexisting under shear cause deviations from a
parabolic flow profile \cite{MM CSA 2003,MM JCP 2003,MM JPCM 2004}. We therefore expect that Eq. ({\ref{eq:convoluted_spectrum})
cannot be used. In fact, the shape of the power spectrum in the crystalline state at low electric field
 strengths is quite different to that of the stationary
state at large fields, where the systems shear melts within a few minutes. This is seen for the lowest applied field strength in
Fig. \ref{fig:11}a, for PnBAPS68 at a particle density of $n = 5.8\;\! \mu\mathrm{m}^{-3}$. A qualitative estimate of the flow
behavior was obtained from microscopy. At low field strength $E$, one observes some kind of double plug flow, lubricated by a
layer crystal region. At a value $E = 22.05$ V/cm, the layer crystal fills the complete cell and the flow profile takes a
parabola-like shape. At $E = 34.6 \;\!$ V/cm and larger, the system is shear-molten at mid-cell height. The system of Fig.
\ref{fig:10}b is at $n  = 26.5\;\! \mu\mathrm{m}^{-3}$ and stays fully crystalline even at the largest applied field strength.
Microscopic observation here reveals a double plug flow in opposing directions. The directly observed flow behavior is mirrored
in the spectral shapes, and none of these spectra can be fitted well by a parabolic flow profile according to Eq.
(\ref{eq:convoluted_spectrum}). Interestingly, this applied also to the systems which were completely shear-molten at mid-cell
height. Height dependent measurements show that this is due to an incomplete shear melting in $y$-direction, altering thus the
aspect.

Clearly, the homogeneity assumption used in the derivation of the scattering theory is valid only locally in (partially)
crystalline systems. and invalidated at each slip line and/or phase boundary, while at the same time the long time
self-diffusion coefficients are expected to differ considerably for different suspension structures  \cite{Simon}. Therefore, Eq.
(\ref{eq:convoluted_spectrum}) has to be applied with great caution in both integral and local measurements. For the low
diffusivity case of crystalline or partially shear-molten states under hydrostatically induced flow through a capillary, the
comparison of local and integral spectra, however, suggests that in a qualitative way the latter can still be interpreted as
velocity distributions and estimates, e.g., of the electrokinetic velocities be obtained \cite{MM JPCM 2004}. Moreover, Fig. 10
has shown that the spectral power does not change upon changing the suspension structure. This allows further quantitative
evaluation of flow profiles, as the intensity at a given frequency is then proportional to the number of scatterers moving with
the corresponding velocity. Hence, within the uncertainty of unknown but presumably small diffusional broadening, quite complex
flow profiles should be analyzable at least on a qualitative level.
In fact, flow profiles consistent with theoretical expectations for multiphase flow have
been extracted from point-by-point measurements of local velocities in pipe flows under hydrostatic pressure differences
\cite{TPSvHMW multiphase flow, Preis flow profiles}.

\section{Conclusions}

Integral heterodyne velocimetry is an established method to measure flow in colloidal systems. It was applied in the past both
in electrokinetic  and rheologic contexts to study electrophoretic mobilities and multi-phase flow, and the theoretical
expressions derived for non-interacting suspensions were applied to estimate velocity distributions and flow profiles. In the
present paper, we went beyond this ad-hoc ansatz and extended the experimental method to super-heterodyning. Moreover, we could
complement this by a derivation of a theoretical expression accounting for structure formation and the unavoidable
polydispersity of real colloidal systems and a careful discussion of the performance and scope of this approximate approach. Our
analysis shows that for small size polydispersity and sufficiently low isothermal compressibility (more precisely: low-$q$
coherent scattering intensity) of the suspension, the signal in the low-$q$ limit should be dominated by incoherent
scattering. Experimentally, a low compressibility was realized by using strongly interacting low-salt suspensions, in the fluid
or crystalline state. In our theoretical considerations, we first assumed a constant particle electrophoretic velocity and a
homogeneous suspension. Experimentally, we studied electrokinetic flow in closed cells yielding parabolic flow profiles. Spatial
variations of the velocity were theoretically accounted for by convoluting a theoretical expression for the particle velocity
distribution with the single-velocity spectral power form. In this way, we obtained an excellent description of the measured power
spectra for fluid-ordered systems.

Under the assumptions made in the scattering theory presented here, the derived expression for the incoherent spectra is
identical in form, but not in value, to that of the non-interacting case. It can be fitted to experimental spectra using four
independent parameters of which two (effective diffusion coefficient and integrated spectral power) are disregarded in the further
evaluation for the electrokinetic velocities. This neglect works well even in the case of crystals and of field-induced
structural changes [26, 27, 28]. The electrophoretic mobility for shear-melting systems was found to be constant, irrespective of
the applied field strength and the stationary system microstructure. This finding is fully consistent with recent simulations
including hydrodynamic interactions \cite{ELS-PRL}. So there already is a practical assessment of this approach for the
complicated case of electrokinetic flow and shear. Inclusion of different experimental situations, like hydrostatic tube flow,
Couette-shear, sedimentation, and even shear-banding flows appears to be rather straightforward. Hence the presented method promises
to become a valuable tool in the studies of colloidal dispersions under flow.

A deeper theoretical understanding, however, remains as a challenge, to be addressed in future work. In particular, the observed
dependencies of the effective diffusion coefficient and the spectral power underline the importance of fluctuations, which have not been
considered so far in the above mentioned evaluations but should be included in
our simple theoretical model. Irrespective of
their origin (shear-induced Taylor dispersion, hydrodynamic coupling by micro-ion currents, polydispersity in the electrokinetic
mobility or sedimentation velocity), it is reasonable to expect that differential motion of neighboring particles will affect the
suspension microstructure and may couple to density fluctuations on the scale of nearest neighbors
distances and larger. So in all these
cases, we expect an enhanced contribution of coherent scattering to the spectra. In particular, an extension of the
theoretical treatment to the coherent dynamic structure factor, $S(\qq,\tau)$, is highly desired, which also includes a worked-out calculation of
its self-part $G(\qq,\tau)$. This should also include strongly interacting suspensions with fully developed fluid microstructure
and polydispersity effects. This, of course is quite demanding, since there is no general expression available to date for the steady-state
dynamic structure factor $S(\qq,t)$ in driven systems covering all cases of driving fields ranging from simple shear over
sedimentation, where we have a relative motion of colloidal particles against the backflowing solvent, to electrokinetic motion
which combines shear flow with particle motion against the solvent. In all three cases the coupling between structure and
diffusivity via direct interactions and hydrodynamic interactions is different, and in any case very complex. A practical aid in
this task could again be given by the experiment, by comparing angle-dependent measurements in the three different situations. This
could clarify whether the changes in the power spectrum are caused by the particle motion relative to the solvent, the shear flow or the applied electric field.
Special interest may further be given to the flow forms in binary mixtures of
low-polydispersity samples. By a suitable choice of
particle material, size and charge, each of the mentioned external perturbations may act differently on the two species and
induce clearly separated signatures in the spectra.

The present paper has reported on an important methodical development, both from the
experimental and theoretical points of view. It sets the stage for progressing from the quantitative characterization of flow in
terms of transport coefficients to a deepened and more general understanding of the complex interplay between particle
interactions, microstructure and dynamics under flow.

\section*{Acknowledgements}
It is a pleasure to thank Nina Lorenz for providing the static
structure factor measurements, and Wolfgang Paul and Mathieu McPhie for helpful
discussions. Financial support of the DFG (SFB TR6, project sections B1 and B2),
and the MWFZ, Mainz is gratefully acknowledged.  \\

\end{document}